\documentclass[10pt,journal,compsoc]{IEEEtran}
\usepackage{ifpdf}

\usepackage{graphicx}
\usepackage{algorithm}
\usepackage{algpseudocode}
\usepackage{booktabs}
\usepackage{subfigure}
\usepackage{amssymb}
\usepackage{multirow}
\usepackage{amsmath}
\usepackage{ragged2e}
\usepackage{array}
\usepackage{caption}

\usepackage{color} 

\ifCLASSOPTIONcompsoc
  \usepackage[nocompress]{cite}
\else
  \usepackage{cite}
\fi
\ifCLASSINFOpdf
\else
\fi
\usepackage{amsmath}
\usepackage{url}

\begin{document}
%

\title{XiHe: A Data-Driven Model for Global Ocean Eddy-Resolving Forecasting}

%
%
%

\author{Xiang Wang, Renzhi Wang, Ningzi Hu, Pinqiang Wang, Peng Huo, Guihua Wang, Huizan Wang, Senzhang Wang, Junxing Zhu, Jianbo Xu, Jun Yin, Senliang Bao, Ciqiang Luo, Ziqing Zu, Yi Han, Weimin Zhang, Kaijun Ren, Kefeng Deng, Junqiang Song
	
\IEEEcompsocitemizethanks{
\IEEEcompsocthanksitem Xiang Wang is with the College of Meteorology and Oceanography, National University of Defense Technology, Changsha 410073, China and with the Department of Atmospheric and Oceanic Sciences, Fudan University, Shanghai 200438, China.
\IEEEcompsocthanksitem Junxing Zhu, Weimin Zhang, Huizan Wang, Yi Han, Jianbo Xu, Pinqiang Wang, Senliang Bao, Ciqiang Luo, Kaijun Ren, and Junqiang Song are with the College of Meteorology and Oceanography, National University of Defense Technology, Changsha 410073, China.
\IEEEcompsocthanksitem Renzhi Wang, Jun Yin and Senzhang Wang are with the School of Computer Science and Engineering, Central South University, Changsha 410083, China.
\IEEEcompsocthanksitem Ningzi Hu is with the College of Oceanography and Space Informatics, China University of Petroleum (East China), Qingdao 266580, China.
\IEEEcompsocthanksitem Peng Huo is with the College of Artificial Intelligence, Tianjin University of Science and Technology, Tianjin 300457, China.
\IEEEcompsocthanksitem Guihua Wang is with the Department of Atmospheric and Oceanic Sciences, Fudan University, Shanghai 200438, China.
\IEEEcompsocthanksitem Ziqing Zu, Key Laboratory of Marine Hazards Forecasting, National Marine Environmental Forecasting Center, Ministry of Natural Resources, Beijing 100081, China.
\IEEEcompsocthanksitem Guihua Wang, Huizan Wang, Senzhang Wang, and Weimin Zhang are the corresponding authors.
}
}

\markboth{Manuscript}
{Shell \MakeLowercase{\textit{et al.}}: Bare Demo of IEEEtran.cls for Computer Society Journals}
%



\IEEEtitleabstractindextext{%
\begin{abstract}
Global ocean forecasting is fundamentally important to support marine activities.
The leading operational Global Ocean Forecasting Systems (GOFSs) use physics-driven numerical forecasting models that solve the partial differential equations with expensive computation.
Recently, specifically in atmosphere weather forecasting, data-driven models have demonstrated significant potential for speeding up environmental forecasting by orders of magnitude, but there is still no data-driven GOFS that matches the forecasting accuracy of the numerical GOFSs.
In this paper, we propose the first data-driven $1/12^{\circ}$ resolution global ocean eddy-resolving forecasting model named \textit{XiHe}, which is established from the 25-year France Mercator Ocean International's daily GLORYS12 reanalysis data. 
\textit{XiHe} is a hierarchical transformer-based framework coupled with two special setups. One is the land-ocean mask mechanism for focusing exclusively on the global ocean circulation. The other is the ocean-specific block for effectively capturing both local ocean information and global teleconnection.
Extensive experiments are conducted under satellite observations, \textit{in situ} observations, and the IV-TT Class 4 evaluation framework of the world's leading operational GOFSs from January 2019 to December 2020. 
The results demonstrate that \textit{XiHe} achieves stronger forecast performance in all testing variables than existing leading operational numerical GOFSs including Mercator Ocean Physical SYstem (PSY4), Global Ice Ocean Prediction System (GIOPS), BLUElink OceanMAPS, and Forecast Ocean Assimilation Model (FOAM). 
Particularly, the accuracy of ocean current forecasting of \textit{XiHe} out to 60 days is even better than that of PSY4 in just 10 days. 
Additionally, \textit{XiHe} is able to forecast the large-scale circulation and the mesoscale eddies. 
Furthermore, it can make a 10-day forecast in only 0.35 seconds, which accelerates the forecast speed by thousands of times compared to the traditional numerical GOFSs. 

\end{abstract}

}

\maketitle

\IEEEdisplaynontitleabstractindextext

%
\IEEEpeerreviewmaketitle

\IEEEraisesectionheading{\section{Introduction}\label{sec:introduction}}
Ocean forecasting is critically important for many marine activities. 
At present, the leading GOFSs (e.g. Mercator Ocean Physical SYstem (PSY4) and Real-Time Ocean Forecast System (RTOFS)) use physics-driven models in fluid mechanics and thermodynamics to predict future ocean motion states and phenomena based on current ocean conditions \cite{francis2020high}. 
The GOFSs adopt numerical methods that rely on supercomputers to solve the partial differential equations of the physical models.
Due to their desirable performance, they are operationally run in different countries worldwide.

However, numerical forecasting methods are usually computationally expensive and slow. 
For example, a single forecasting simulation in the numerical GOFSs may take hours on a supercomputer with hundreds of computational nodes~\cite{tonani2015status}. 
Besides, improving the forecasting accuracy of these methods is exceedingly challenging because they heavily rely on the human cognitive abilities in understanding the physical laws of the ocean environment \cite{Peter2015Nature}.

With the recent advances of Artificial Intelligence (AI) techniques, deep learning methods have been widely applied in various prediction/forecasting tasks of different fields and achieved great success. 
Particularly, some data-driven AI models have shown the potential in atmosphere weather forecasting like \textit{Pangu-Weather}~\cite{pangu2023nature} and \textit{GraphCast}~\cite{GraphCast2023Science}.
They have achieved comparable or even better prediction results in global medium-range weather forecasting than current leading numerical weather prediction (NWP) methods\cite{FourCastNet2022,pangu2023nature,GraphCast2023Science,ClimaX2023,fengewu2023,fuxi2023}.
One significant advantage of data-driven models is that they can make the forecasting thousands or even tens of thousands of times faster than NWP methods \cite{pangu2023nature}. 
Furthermore, they can automatically learn the spatial-temporal relationships from massive meteorological data, and effectively capture the rules of weather changing, without introducing the prior knowledge of physics mechanisms.

Although data-driven models have achieved promising results in atmosphere weather forecasting, how to build a more accurate and efficient data-driven ocean forecasting model remains an open research issue due to the following two challenges. 
First, different from being interconnected in the atmospheric space, the ocean is divided into several relatively independent regions with specific water mass properties due to the existence of continents and islands. 
Therefore, it is more difficult to automatically learn the internal changing rules and mutual influence mechanisms of different ocean regions. 
Second, the horizontal resolution of the typical operational GOFSs ($1/12^{\circ}$) is much higher than that in the data-driven global weather prediction systems ($1/4^{\circ}$) \cite{FourCastNet2022,pangu2023nature,GraphCast2023Science,fengewu2023,fuxi2023}.
Since training a deep learning-based global medium-range weather prediction model requires extensive computing resources and time, there is a need to develop a more efficient data-driven global ocean forecasting model with $1/12^{\circ}$ horizontal resolution. 

To address the above challenges, we propose the first data-driven $1/12^{\circ}$ resolution global ocean eddy-resolving forecasting model \textit{XiHe}, which adopts a hierarchical transformer framework. 
Specifically, we introduce the land-ocean mask mechanism to explicitly mask out the continent and island regions, which assists \textit{XiHe} to focus more on learning the inherent changes of the global ocean circulation. 
We also propose an ocean-specific block in the framework for effectively capturing both local and global ocean physical information. 
The ocean-specific block contains several local spatial information extraction (SIE) modules and a global SIE module. 
The local SIE module applies the window-attention mechanism to capture the spatial dependencies in local oceanic regions. 
The global SIE module aims to capture the global oceanic properties and the interactions among the local regions based on the cross-attention mechanism. 
Eventually, the whole architecture of \textit{XiHe} is constructed by stacking several ocean-specific blocks. 
As the ocean-specific block has linear computational complexity to the data size, \textit{XiHe} can handle high-resolution input data efficiently. 
Extensive experiments conducted under satellite observations, \textit{in situ} observations, and the IV-TT Class 4 evaluation framework of the world's leading operational GOFSs, verify the effectiveness and efficiency of \textit{XiHe} in global ocean forecasting.

We summarize our main contributions as follows. 
\begin{enumerate}
\item We propose the first data-driven $1/12^{\circ}$ resolution global ocean eddy-resolving forecasting model \textit{XiHe}, which achieves stronger forecast performance in all testing variables than the compared leading operational numerical GOFSs. Moreover, \textit{XiHe} can make a forecast in only 0.35 seconds (1000x faster than the numerical GOFSs) averagely with one GPU. 
\item We introduce an ocean-land masking mechanism to exclude the impact of land, thereby focusing exclusively on learning global ocean information. We also design a novel ocean-specific block containing local and global SIE modules to capture the inherent oceanic spatial information. 
\item Extensive experiments on the authoritative IV-TT Class 4 evaluation framework verify that the accuracy of ocean current forecasting of \textit{XiHe} out to 60 days is even better than that of PSY4 in 10 days. 
\item Experimental evaluations based on satellite and \textit{in situ} observations demonstrate that \textit{XiHe} is able to forecast the large-scale circulation and the mesoscale eddies.
\end{enumerate}

The remainder of this paper is organized as follows. 
Section \ref{sec:Relatedworks} briefly reviews the previous related works.  
Section \ref{sec:Dataset} introduces the experimental dataset. 
Section \ref{sec:Method} illustrates the framework of the data-driven model \textit{XiHe}. 
Section \ref{sec:Experiments} conducts experiments to compare the performances of \textit{XiHe} with existing leading numerical GOFSs, and proves its advantages on global ocean forecasting. 
Section \ref{sec:conclusion} draws conclusions and discussions.

\section{Related Work} \label{sec:Relatedworks}

Ocean forecasting aims to provide accurate and timely forecasts of the future ocean conditions based on current and historical ocean observations.
Existing methods can generally be categorized into the physics-driven conventional NWP methods and the data-driven methods. 
The NWP methods have a long research history and have been widely applied, while the data-driven methods have emerged recently and developed rapidly. 

\subsection{Physics-Driven Methods}

With the advancements in physical theory, numerical model, and supercomputing, physics-driven numerical methods have become the leading operational methods for ocean forecasting in most countries~\cite{tonani2015status}.
Existing ocean forecasting systems generally use an ocean numerical model as the dynamic framework, and integrate the near real-time high-quality observations into the model by data assimilation to realize ocean forecasting~\cite{tonani2015status,liu2023global}.

The major maritime nations in the world have paid considerable attention to ocean forecasting, and a series of ocean numerical forecasting models and operational forecasting systems have been established. 
The leading numerical forecasting models include the HYbrid Coordinate Ocean Model (HYCOM)~\cite {HYCOM}, Nucleus for European Modelling of the Ocean (NEMO)~\cite{NEMO}, Modular Ocean Model (MOM)~\cite{MOM} and so on. 
The leading ocean operational forecast systems include
BLUElinK OceanMAPS (BLK) from the Australian Bureau of Meteorology~\cite{BLK}, 
the Global Ice Ocean Prediction System (GIOPS) from the Canadian Meteorological Center~\cite{GIOPS}, 
Forecast Ocean Assimilation Model (FOAM) from the UK Meteorological Office~\cite{FOAM},
Mercator Ocean Physical SYstem (PSY4) from the France Mercator Océan International~\cite{PSY}, 
Real-Time Ocean Forecast System (RTOFS) from the National Centers for Environmental Prediction~\cite{RTOFS} and so on.

Until now, numerical ocean forecasting methods have contributed overall the highest prediction accuracy, but they still face two major limitations. 
Firstly, they are usually computationally expensive and slow due to the high complexity of solving the physical partial differential equations~\cite {pangu2023nature}.
Secondly, the improvement of the forecasting accuracy is exceedingly challenging due to their limitation of human cognitive abilities in understanding the physical laws of the ocean environment~\cite{Peter2015Nature}.

\subsection{Data-Driven Methods}

With the fast development of AI techniques, data-driven AI models have achieved great success in atmosphere weather forecasting~\cite{GraphCast2023Science}. 
Early data-driven methods are primarily used for precipitation forecasting from radar images, where traditional NWP methods are relatively weak~\cite{DGMR2021nature, espeholt2022Nature_comm, NowcastNet2023nature}.
Recent data-driven models focus on global medium-range atmosphere weather forecasting. 
Some data-driven atmosphere weather forecasting models like FourCastNet \cite{FourCastNet2022}, Pangu-Weather \cite{pangu2023nature}, GraphCast \cite{GraphCast2023Science}, FengWu \cite{fengewu2023}, and FuXi \cite{fuxi2023} achieve comparable or even better accuracy in some weather variables when compared to Integrated Forecasting System (IFS), which is the top NWP system. 
Furthermore, these data-driven models are orders of magnitude faster than NWP methods. 
For example, GraphCast can make 10-day forecasts under one minute and outperforms IFS on 90\% of verification targets~\cite{GraphCast2023Science}. 
Pangu-Weather surpasses IFS in all tested variables, while being over 10,000-times faster~\cite{pangu2023nature}.

The data-driven methods have also shown promising performance in ocean forecasting tasks such as forecasting the sea surface field of individual variables (e.g. Satellite Sea Surface Temperature (SST), Sea Surface Height (SSH), wave height)~\cite{Zheng2020Purely,quach2020deep}, ocean phenomena on the seasonal or interannual scale (e.g. Indian Ocean Dipole (IOD), El Niño Southern Oscillation (ENSO))~\cite {ham2019deep,zhou2022hybrid,zhou2023self}, and global multi-variables ocean forecasting~\cite {AI-GOMS2023}.
For example, Zheng et al. ~\cite {Zheng2020Purely} designed a satellite data–driven deep learning model for forecasting the SST field.
Ham et al. ~\cite {ham2019deep} used transfer learning to train a Convolutional Neural Network(CNN) model for skillful ENSO prediction. 
Zhou and Zhang ~\cite {zhou2023self} constructed a transformer-based neural network named 3D-Geoformer for spatiotemporal multivariate predictions of ENSO. 
Xiong et al. ~\cite {AI-GOMS2023} built a global ocean modeling System AI-GOMS for multi-variables ocean forecasting. AI-GOMS adopted the Fourier-based Masked Auto-encoder structure as the backbone model for basic ocean variables prediction.
Their global ocean forecast is at $1/4^{\circ}$ spatial resolution and their region forecast is at $1/12^{\circ}$ spatial resolution.
However, it is difficult for AI-GOMS to perform global ocean forecasting at $1/12^{\circ}$ spatial resolution in an acceptable training time, due to its $O(NlogN)$ ($N$ is the data size) computational complexity. 
Furthermore, its forecast results are not compared with any numerical GOFSs, such as GIOPS, FOAM, BLUElinK OceanMAPS, PSY4, and RTOFS. 

To sum up, recently there have been several data-driven models in global atmosphere weather forecasting, which achieved comparable or even better prediction performance than NWP methods. 
Furthermore, data-driven methods have also shown the potential in accelerating numerical ocean forecasting by orders of magnitude, but there is still a significant gap in the forecasting accuracy compared with the numerical operational GOFSs at $1/12^{\circ}$ resolution.



\section{Dataset}\label{sec:Dataset}
\subsection{Training Dataset}
The main training data of \textit{XiHe} is the GLORYS12 reanalysis which is designed and implemented in the framework of the Copernicus Marine Environment Monitoring Service (CMEMS)~\cite{GLORYS12}. The GLORYS12 is a global reanalysis at $1/12^{\circ}$ horizontal resolution with 50 vertical levels. In this dataset, along track altimeter Sea Level Anomaly (SLA), SST, and Sea Ice Concentration (SIC), as well as \textit{in situ} Temperature and Salinity (T/S) profiles are jointly assimilated. The GLORYS12 dataset we utilized spans from Jan. 1993 to Dec. 2020. GLORYS12 is one of the state-of-the-art reanalyses~\cite{fu2023global}. Due to its high resolution, time continuity, long coverage, and excellent quality, we choose it as the main training dataset. 

Considering the impact of the sea surface wind on ocean dynamics, we incorporate the 10m u-component of wind (U10) and 10m v-component of wind (V10) into the training data of \textit{XiHe}. The wind field data for training is from the Fifth Generation Global Atmospheric Reanalysis data (ERA5) produced by the European Centre for Medium-Range Weather Forecasts (ECMWF) \cite{2021The}. Their spatial resolution is 0.25° with 137 model levels. In order to standardize the data dimensions, we interpolated the wind data to the same spatial and temporal resolution as the GLORYS12 reanalysis data. 

To improve the forecasting accuracy of SST, we introduce the satellite SST data from the Operational Sea Surface Temperature and Ice Analysis (OSTIA) to the training data of \textit{XiHe}. OSTIA is published by Met Office based on data provided by GHRSST-PP, which uses AATSR counts, SEVIR/data, AVHRR data, AMSR data, TMI data, and measured data, with a spatial resolution of approximately 5km and a grid spacing of $1/20^{\circ}$. OSTIA provides near-real-time high spatial resolution SST data \cite{DONLON2012140}. We also interpolated SST data to the same spatial resolution as the GLORYS12 reanalysis data.

\subsection{Evaluation Dataset}\label{sec:EvaDataset}
\textit{XiHe} is primarily evaluated based on the GODAE OceanView Inter-comparison and Validation Task Team (IV-TT) Class 4 framework, which provides observation datasets from the Argo, the USGODAE server, and the Jason-1, Jason-2, and Envisat CLS AVISO level 3 satellite altimeters~\cite{IV-TT}. 
(1) Argo is an international global observational array of nearly 4,000 autonomous robotic profiling floats, each measuring ocean temperature and salinity from sea surface to about 2,000m on nominal 10-day cycles. 
(2) \textit{In situ} drifting buoys in USGODAE provide SST and near-surface currents, with most buoy sites distributed between 60°N and 60°S, and the number of buoy sites varies from 20,000 to 30,000 every day. 
(3) The Jason-1, Jason-2, and Envisat CLS AVISO level 3 satellite altimeters provide daily along-track SLA data. 

In addition, we evaluate the ocean temperature and salinity based on the Tropical Atmosphere Ocean (TAO) project, the Research Moored Array for African–Asian–Australian Monsoon Analysis and Prediction (RAMA), and the Prediction and Research Moored Array in the Tropical Atlantic (PIRATA), which are three significant oceanographic research programs. 
Each focuses on different parts of the world’s oceans and plays a vital role in monitoring and understanding oceanic and atmospheric processes. 
The TAO project concentrates on the Pacific Ocean, the RAMA project focuses on the Indian Ocean, and the PIRATA project is dedicated to the Tropical Atlantic Ocean, comprises moored buoys that provide critical real-time data on oceanic and atmospheric conditions. 


The CMEMS provides global ocean gridded L4 sea surface heights and derived variables reprocessed data with a spatial resolution of $1/4^{\circ}$ from 1993 to 2023~\cite{os-15-1207-2019}. 
These data's product ID is SEALEVEL\_GLO\_PHY\_L4\_MY\_008\_047. 
The product includes daily SLA data, absolute dynamic topography, geostrophic currents and so on. 
The daily SLA data is estimated by optimal interpolation, merging the L3 along-track measurement from the different altimeter missions available. 
The geostrophic currents data is a derived product of altimeter satellite gridded sea surface heights. 
According to the quality information document, the estimated errors on geostrophic currents range between 9 and 16 cm/s. 
We employ the data to evaluate the performance of \textit{XiHe} on the surface currents and ocean mesoscale eddies.

\section{Method}\label{sec:Method}
\subsection{Problem Statement}
In this work, we focus on predicting 5 ocean variables including ocean temperature, salinity, zonal and meridional components of ocean currents with 23 layers (\textit{i.e.}, 0.49m, 2.65m, 5.08m, 7.93m, 11.41m, 15.81m, 21.60m, 29.44m, 40.34m, 55.76m, 77.85m, 92.32m, 109.73m, 130.67m, 155.85m, 186.13m, 222.48m, 266.04m, 318.13m, 380.21m, 453.94m, 541.09m and 643.57m in depth), and SSH. 

In this work, the input variables of our \textit{XiHe} model include 2 sea surface variables (sea surface temperature and sea surface height), 4 ocean variables with 23 layers (ocean temperature, sea salinity, zonal and meridional component of ocean currents), and 2 atmospheric variables of sea surface (zonal and meridional component of sea surface wind), a total of 96 variables. 
The sea surface temperature data is from OSTIA and data of other ocean variables comes from the GLORYS12 global reanalysis dataset. 
The GLORYS12 global reanalysis data has a spatial resolution of $1/12^{\circ}$ ($4320\times2041$ longitude-latitude grid points) and a temporal resolution of one day. 
Two atmospheric variables (zonal and meridional components of sea surface wind) are added to help learn ocean dynamics raised by atmosphere forcing, which are from the ERA5 reanalysis dataset from ECMWF. 
We stack these variables to generate the input data tensor $\mathbf{X}^{t}\in \mathbb{R}^{W \times H \times C_{\text {in}}}$, where $W$ and $H$ denote the number of grid points in longitude and latitude directions, respectively. 
$C_{\text {in}}$ denotes the number of input variables. 
With the $1/12^{\circ}$ spatial resolution and 96 input variables, we have $W=4320$, $H=2041$ , and $C_{in}=96$. 
Given the input data tensor $\mathbf{X}^{t}$ corresponding to the current time step, the objective of \textit{XiHe} is to output the $K$-step-ahead forecasts $\hat{\mathbf{X}}^{t+1: t+K}=\left(\hat{\mathbf{X}}^{t+1}, \ldots, \hat{\mathbf{X}}^{t+K}\right)$, $\hat{\mathbf{X}}^{t+\tau} \in \mathbb{R}^{W \times H \times C_{\text {out}}}$,$\tau=1,2,...K$. 
The data dimensions of the output $\hat{X}^{t+\tau}$ are $4320\times2041\times94$ (2 atmospheric variables are excluded). The forecasting task can be formulated by:
\begin{equation}
    \hat{\mathbf{X}}^{t+\tau}=\mathcal{F}\left(\mathbf{X}^{t}, \theta\right),\tau=1,2,...,K
\end{equation}
where $\mathcal{F}(\cdot
)$ is our data-driven model \textit{XiHe} and $\theta$ denotes the model parameters.

\subsection{Framework Overview} 
\begin{figure*}
    \centering
    \includegraphics[width=0.98\textwidth]{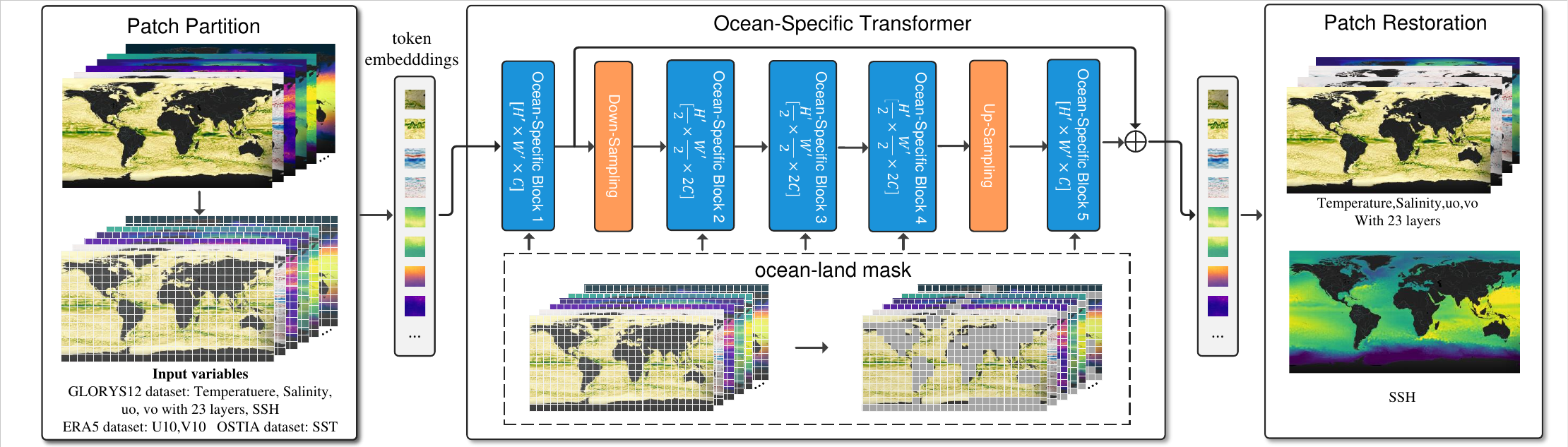}
    \caption{The overall framework of \textit{XiHe}, which contains the Patch Partition module, Ocean-Specific Transformer and the Patch Restoration module. The Ocean-Specific Transformer consists of 5 ocean-specific blocks, a down-sampling block and an up-sampling block. The 1st and the 5th ocean-specific blocks consist of a local SIE module and a global SIE module, and the 2nd to the 4th ocean-specific blocks consist of 2 consecutive local SIE modules and a global SIE module. The ocean-land masking mechanism is designed to make the model better focus on the ocean dynamics.}
\label{framework}
\end{figure*}

As illustrated in Fig.~\ref{framework}, \textit{XiHe} is a hierarchical transformer-based machine learning framework, which contains three components: the patch partition module, ocean-specific transformer, and the patch restoration module.
As shown in the left part of Fig.~\ref{framework}, the input data tensor $\mathbf{X}^{t}$ is firstly transformed into a $C$-dimension latent space through the patch partition module to obtain $C$-dimension token embeddings. 
Next, as shown in the middle part of Fig.~\ref{framework}, we stack 5 ocean-specific blocks in the Ocean-Specific Transformer to extract the local and global ocean information from the token embeddings. 
Additionally, we introduce an ocean-land masking mechanism in the ocean-specific block.
Intuitively, the ocean-land masking mechanism functions as the prior knowledge that the earth's ocean is divided into several relatively independent regions by the continents and islands. 
Therefore, it can exclude the influence from irrelevant lands and help \textit{XiHe} better focus on learning ocean dynamics.
Since the proportion of ocean and land on Earth is 70.8\% and 29.2\% respectively, \textit{XiHe} can reduce a significant amount of computational cost by the masking mechanism.
A down-sampling block and an up-sampling block are also included to produce hierarchical spatial representations. 
Finally, a patch restoration module is used to generate the forecast results as shown in the right part of Fig.~\ref{framework}. 
The details of each component in \textit{XiHe} will be introduced in the following subsections.
 
\subsection{Patch Partition and Patch Restoration}
To reduce the spatial resolution and accelerate the training process, we follow the patch partition scheme in the standard vision transformer\cite{VIT2021} to divide the input data tensor $\mathbf{X}^{t}\in \mathbb{R}^{W \times H \times C_{\text {in}}}$ to ${W}^{'} \times {H}^{'}$ patches with non-overlapping windows. 
It adopts a 2-dimension (2D) convolution layer with the kernel size $p$. 
The stride of the sliding windows is the same as the kernel size. 
All land grids and missing values are filled with the value of zero.
Additionally, to make sure that the data size can be divisible by the patch size $p$, the zero-value padding is added to the input data. 
The patch partition results in a feature map with the shape of ${W}^{'} \times {H}^{'} \times C$ where ${W}^{'}=\frac{W}{p}$, ${H}^{'}=\frac{H}{p}$, and each patch token is a $C$-dimension vector.
After patch partition, a layer normalization (LayerNorm)\cite{layernorm} is employed to improve training stability.

The patch restoration is the opposite operation of the patch partition, which scales the feature map with the shape of  ${W}^{'} \times {H}^{'} \times C$ up to $W \times H \times C_{\text {out}}$ using a 2D transposed convolution layer with a kernel size of $p$ and a stride of $p$.

\subsection{Ocean-specific Transformer}
\begin{figure}
\centering
\includegraphics[width=0.48\textwidth]{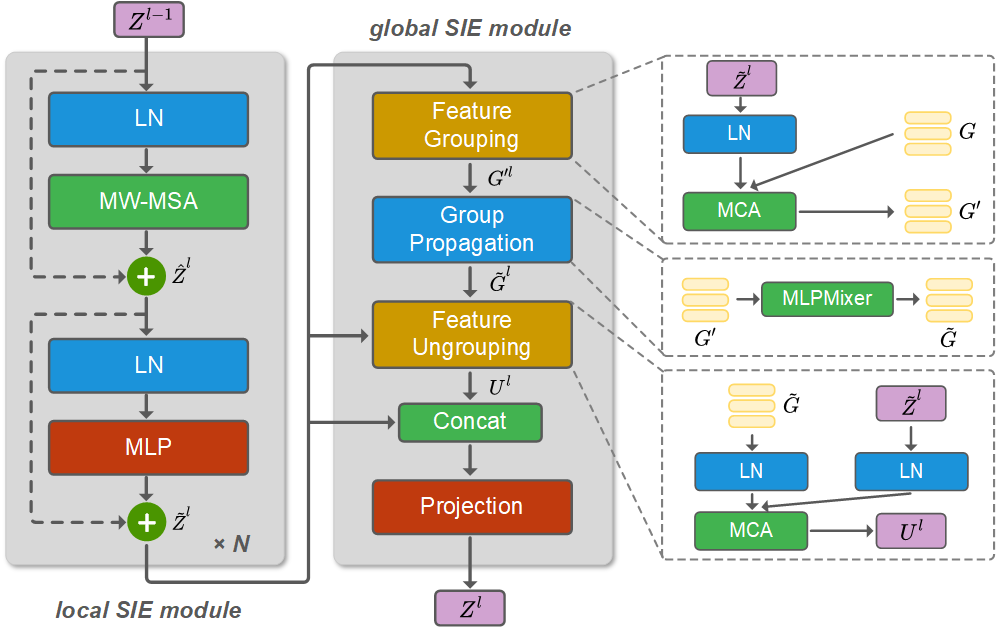}
\caption{Illustration of the ocean-specific block, which contains several consecutive local SIE modules and a global SIE module. We introduce the ocean-land masking mechanism to the W-MSA of the local SIE module and the feature grouping stage of the global SIE module. In this figure, we only give one local SIE module for example.} 
\label{OceanSpecifiedBlock}
\end{figure}

The ocean-specific Transformer consists of 5 ocean-specific blocks, a down-sampling block, and an up-sampling block. 
As illustrated in Fig.~\ref{OceanSpecifiedBlock}, an ocean-specific block contains several consecutive local spatial information extraction (SIE) modules and a global spatial information extraction module. 
Especially, The 1st and the 5th ocean-specific blocks in Fig.~\ref{framework} consist of a local SIE module and a global SIE module, and the 2nd to the 4th ocean-specific blocks consist of 2 consecutive local SIE modules and a global SIE module. 
The input data size remains unchanged in the ocean-specific block. 
To produce a hierarchical representation and model at different spatial scales, the down-sampling block is adopted to reduce the number of patches. 
Specifically, the down-sampling block follows the patch merging implementation in Swin Transformer\cite{Swin2021}. 
It merges $2 \times 2$ neighboring patches into one, resulting in a feature map with the shape of $\frac{{W}^{'}}{2} \times \frac{{H}^{'}}{2} \times 4C$. 
A linear layer follows to reduce the dimension from $4C$ to $2C$. 
The feature map through down-sampling will sequentially pass through the next three ocean-specific blocks. 
The up-sampling block restores the data size back to ${W}^{'} \times {H}^{'} \times C$ by applying the reverse operation of down-sampling. 
Furthermore, a skip connection is adopted to concatenate the output of the 1st and 5th ocean-specific blocks. 
Next, we will describe the technical details of the ocean-specific block.

\subsubsection{Local SIE module}
Recently, owing to the strong ability of the global self-attention mechanism to model global semantic information, Vision Transformer (ViT) \cite{VIT2021} and its variants \cite{Swin2021,crossvit2021,DeiT2022} have achieved great success in various computer vision tasks. 
Nevertheless, the global self-attention mechanism is infeasible for high-resolution data for the reason that its computational complexity is quadratic to the data size. 
To improve the efficiency, we follow the window-attention mechanism in Swin Transformer\cite{Swin2021} to compute multi-head self-attention within local windows in the local SIE module. 
As illustrated in Fig.~\ref{OceanSpecifiedBlock}, the local SIE module is computed as follows:

\begin{align}
{\hat{\mathbf{Z}}}^{l}&=\mathrm{W}\mbox{-}\mathrm{MSA}\left(\mathrm{LN}\left({{\mathbf{Z}}}^{l-1}\right)\right)+{{\mathbf{Z}}}^{l-1},\\
{\tilde{\mathbf{Z}}}^{l}&=\mathrm{MLP}\left(\mathrm{LN}\left({\hat{\mathbf{Z}}}^{l}\right)\right)+{\hat{\mathbf{Z}}}^{l},
\end{align}
where W-MSA denotes the Window-based Multi-head Self-Attention. ${\hat{\mathbf{Z}}}^{l}$ and ${\tilde{\mathbf{Z}}}^{l}$ represent the output feature maps from the W-MSA module and the MultiLayer Perceptron (MLP) module. $LN$ denotes the LayerNorm layer. In the W-MSA module, $\mathbf{Z}^{l-1}$ (the feature map from the last layer) is firstly partitioned by non-overlapping windows of size $M \times M$, each of which contains $M \times M$ patches. Then multi-head self-attention mechanism is applied within each window in parallel, which has linear computational complexity with respect to the data size. The self-attention used inside the window can be formulated as Eq. (\ref{eqaAttention}).
\begin{equation}
  \operatorname{Attention}(\mathbf{Q}, \mathbf{K}, \mathbf{V})=\operatorname{SoftMax}\left(\mathbf{Q} \mathbf{K}^{\top} / \sqrt{D}+\mathbf{B}\right) \mathbf{V},
  \label{eqaAttention}
\end{equation}
where $\mathbf{Q}$, $\mathbf{K}$, and $\mathbf{V}$ are query, key, and value vectors, ${D}$ is the feature dimension and $\mathbf{B}$ denotes the relative position bias.

\subsubsection{Global SIE module} 
The local SIE module ignores the connections between different windows, which hinders the exchange of long-range information within the global ocean. Due to the phenomenon of teleconnection in the ocean system, it is possible to achieve high correlations between different areas that are far away from each other (e.g., ocean teleconnection). 
To address this issue, we introduce a collection of group vectors to learn the global dynamics of the ocean system inspired by Group Propagation Vision Transformer (GPViT) \cite{GPViT2023}. 
Intuitively, characteristics exhibited in different ocean regions may be similar. 
For example, due to the influence of solar radiation with latitude, the sea surface temperatures are lower in the Antarctic and Arctic regions, while they are higher in the areas near the equator. 
We can view the patches with similar characteristics as a group. 
And each of the group vectors represents a group of patches. 
The global SIE module mainly consists of three stages as shown in Fig.~\ref{OceanSpecifiedBlock}: feature grouping, group propagation, and feature ungrouping. 

\textbf{Feature grouping.} In the stage of feature grouping, we first initialize a set of group vectors $\mathbf{G}^l$ randomly. 
The number of group vectors is a model hyper-parameter. 
Then we perform a Multi-head Cross-Attention (MCA) operation to update group vectors, in which the queries are learnable group vectors $\mathbf{G}^l$, and the keys and values are the feature map ${\mathbf{\tilde{Z}}}^{l}$ computed by the preceding local SIE modules as Eq. (\ref{eqaMCA}):
\begin{equation}
  {\mathbf{G}^{\prime}}^{l}=\operatorname{Concat}_{\{h\}}\left(\operatorname{Attention}\left(\mathbf{W}_{h}^{Q} \mathbf{G}_{h}^{l}, \mathbf{W}_{h}^{K} \mathbf{\tilde{Z}}^{l}_{h}, \mathbf{W}_{h}^{V} \mathbf{\tilde{Z}}^{l}_{h}\right)\right),
  \label{eqaMCA}
\end{equation}
where $h$ denotes the $h$-th head, $\mathbf{W}_{h}^{Q}$, $\mathbf{W}_{h}^{K}$ and $\mathbf{W}_{h}^{V}$ are projection matrices for the query, key and value vectors, respectively. Attention denotes the standard attention mechanism in the vanilla Transformer\cite{vaswani2017attention} and ${\mathbf{G}^{\prime}}^{l}$ denotes the updated group vectors. 
During the feature grouping, the ocean representations provided by the local SIE module are clustered into several groups which appear as the prominent ocean patterns, according to their relevance with the group vectors. 
Based on these local ocean representations, the group vectors that summarize the local ocean information are updated preliminarily.
As the number of group vectors is much smaller than the number of patches, the feature grouping has a linear computational complexity with the fixed number of group vectors and the global information can be propagated much faster. 

\textbf{Group propagation.} Next in the stage of group propagation, MLPMixer\cite{MLPMixer2021} is applied to propagate global information among group vectors and fuse them. 
Throughout the group propagation, the group vectors exchange information with each other and thus succeed in depicting the global ocean dynamics from a comprehensive perspective.
The process of MLPMixer can be written as follows:
\begin{align}
    &\hat{\mathbf{G}}^{l} = {\mathbf{G}^{\prime}}^{l} +\mathrm{MLP}_{1}\left(\text{LN}{({\mathbf{G}^{\prime}}^{l})}^{T}\right)^{T},\\
   & \mathbf{\tilde{G}}^{l} = {\hat{\mathbf{G}}}^{l} + \mathrm{MLP}_{2}\left(\text{LN}{(\hat{\mathbf{G}}^{l})}\right),
\end{align}
where ${\mathbf{G}^{\prime}}^{l}$ are group vectors computed in the stage of feature grouping, and $\tilde{\mathbf{G}}^{l}$ are fused group vectors. 

\textbf{Feature ungrouping.} After group propagation, the global information is returned to patch tokens through the stage of feature ungrouping. 
In this stage, the feature map computed by the preceding local SIE module ${\tilde{\mathbf{Z}}}^{l}$ attends to the updated group vectors $\tilde{G}$ via multi-head cross-attention operation as follows:
\begin{equation}
  \mathbf{U}^l=\operatorname{Concat}_{\{h\}}\left(\operatorname{Attention}\left(\tilde{\mathbf{W}}_{h}^{Q} \tilde{\mathbf{Z}}^{l}_{h}, \tilde{\mathbf{W}}_{h}^{K} \tilde{\mathbf{G}}^{l}_{h}, \tilde{\mathbf{W}}_{h}^{V} \tilde{\mathbf{G}}^{l}_{h}\right)\right),
  \label{eqa 1}
\end{equation}
where $h$ denotes the $h$-th head, $\tilde{\mathbf{W}}_{h}^{Q}$, $\tilde{\mathbf{W}}_{h}^{K}$ and $\tilde{\mathbf{W}}_{h}^{V}$ are projection matrices for the query, key and value vectors, respectively. $\mathbf{U}^l$ is then concatenated with ${\tilde{\mathbf{Z}}}^{l}$. The projection layer follows to  transform the concatenated feature map $[\mathbf{U}^l,\tilde{\mathbf{Z}}^{l}]$ to the original dimensions and output the feature map $\mathbf{Z}^{l}$. 
Through the global SIE module, the model grasps the general dynamics of the ocean system.

\subsection{The Ocean-land Masking Mechanism} 
Different from meteorological problems, the ocean environment contains a significant amount of landmasses. 
Though we have filled the land areas with zero-values during data preprocessing, these regions still participate in the calculations during the model training process, which disperses the model's attention to the dynamics of the ocean regions. 
To address this problem, inspired by Swin Transformer\cite{Swin2021}, we propose the ocean-land masking mechanism in order to mask out the land areas when computing attention. 
Firstly, we generate the ocean-land mask matrix $\mathbf{M}$ according to the sea surface data. 
$\mathbf{M}$ is in the shape of $2041\times4320$, in which the value of zero means the grid point is in land areas while the value of one means the grid point is in ocean areas. 
We apply patch partition to $\mathbf{M}$ and if all the grid points in a patch are land areas, we assign the value of zero for this patch otherwise one. 
We follow the rule to generate coarser resolution masks $\mathbf{M_1}\in \mathbb{R}^{{W}^{'} \times {H}^{'} }$ and $\mathbf{M_2}\in \mathbb{R}^{\frac{{W}^{'}}{2} \times \frac{{H}^{'}}{2} }$. 
As shown in the lower middle part of Fig.~\ref{framework}, we mask out the patches for land areas according to $\mathbf{M_1}$ and $\mathbf{M_2}$ so that the attention scores for land areas are assigned to the value of zero in the local SIE module. $\mathbf{M_1}$ is used for the 1st and the 5th ocean-specific blocks and $\mathbf{M_2}$ is used for the 2nd to 4th ocean-specific blocks. Additionally, to further reduce the computational cost, the patches for land areas are excluded from the multi-head cross-attention computation in the stage of feature grouping in the global SIE module.

\subsection{Optimization Details}
\textit{XiHe} generates the predictions in a non-auto-regressive style, which means that it generates the predictions directly from the input data. 
The loss function used is the Mean Square Error (MSE) loss. The MSE loss is defined as Eq. (\ref{mseloss}),
\begin{equation}
    \mathcal{L}_{\mathrm{MSE}}=\frac{1}{W\times H\times C_{out}}\sum_{i=1}^{W}\sum_{j=1}^{H}\sum_{c=1}^{C_{out}}\left(\hat{\mathbf{X}}_{i,j,c}^{t+\tau}-\mathbf{X}_{i,j,c}^{t+\tau}\right)^2,
    \label{mseloss}
\end{equation}
where $C_{out}$, $H$, and $W$ are the number of output variables and the number of grid points in latitude and longitude directions, respectively. $c$, $i$, and $j$ are the indices of output variables, longitudinal grids, and latitudinal grids, respectively. $\hat{\mathbf{X}}_{i,j,c}^{t+\tau}$ and $\mathbf{X}_{i,j,c}^{t+\tau}$ are predicted values and ground truth for some target variables and locations at the time step of $t+\tau$. \textit{XiHe} is implemented by the Pytorch framework. Each model is trained for 50 epochs using the batch size of 1 on each GPU. We use the AdamW\cite{adam,decoupled} optimizer with the parameters $\beta_1=0.9$, $\beta_2=0.95$ and a learning rate $5e^{-5}$. Additionally, a scheduled DropPath with a drop ratio of 0.2 is applied to avoid over-fitting.

\section{Experimental Evaluation}~\label{sec:Experiments}

The forecasting capacity of the \textit{XiHe} is evaluated based on satellite observations, \textit{in situ} observations, and the IV-TT Class 4 evaluation framework of the world's leading operational GOFSs. 
Note that \textit{XiHe} forecasts the oceanic variables in a $1/12^{\circ}$ spatial resolution (grid format of 2041$\times$4320), which is the same as the GLORYS12 reanalysis data.

\subsection{Overall Evaluation based on IV-TT Class 4 Framework}

\subsubsection{The IV-TT Class 4 Framework}
We evaluated the performances of \textit{XiHe} under the GODAE OceanView IV-TT Class 4 framework, which is designed to evaluate the world's leading operational ocean forecasting systems like FOAM, PSY4, GIOPS, BLUElink OceanMAPS, and RTOFS~\cite{IV-TT}. 
The IV-TT Class 4 framework focuses on the assessment of the forecast performance and skill, using available ocean observations including \textit{in situ} drifting buoys and satellite data in near real-time. 
Within the IV-TT Class 4 framework, each forecasting system interpolates its forecast fields, including ocean temperature profile, ocean salinity profile, SLA, etc., to the observation locations. 
Then different forecasting systems can be evaluated using the observations and compared with each other. 
The framework utilizes ocean Temperature and Salinity  observations that are sourced from Argo profiles. Additionally, SLA was measured by satellite altimeters~\cite{IV-TT}.

There are five statistical metrics in the IV-TT Class 4 framework, including Bias, root mean square error (RMSE), ACC, and two skill scores constructed using persistence and climatology fields~\cite{IV-TT}. 
The RMSE, ACC, and Bias are used in this paper. 
According to the evaluation of the IV-TT framework, we first interpolate the grid forecast results into the IV-TT observations and then calculate the evaluation metrics~\cite{IV-TT}. 
For example, the RMSE and ACC are defined in Eq. (\ref{eq:rmse}) and Eq. (\ref{eq:acor}). 
\begin{equation}
    {\rm RMSE}=\sqrt{\frac{1}{N}\sum_{i=1}^{N}(\mathbf{F}_i-\mathbf{O}_i)^2}\label{eq:rmse},
\end{equation}
\begin{equation}
    {\rm ACC}=\frac{\sum_{i=1}^{N}(\mathbf{F_i}-\mathbf{C_i})(\mathbf{O_i}-\mathbf{C_i})}{\sqrt{\sum_{i=1}^{N}(\mathbf{F_i}-\mathbf{C_i})^2} \sqrt{\sum_{i=1}^{N}(\mathbf{O_i}-\mathbf{C_i})^2}\label{eq:acor}},
\end{equation}
where $N$ denotes the number of the observational points in evaluation, $\mathbf{F}_i$ denotes the forecast value of the $ith$ point, $\mathbf{O}_i$ denotes the observational value of the $ith$ point, and $\mathbf{C_i}$ is the corresponding climatology of the $ith$ point provided by the IV-TT Class 4 framework.

We compare \textit{XiHe} with the four state-of-the-art global ocean forecasting systems, including PSY4 \cite{PSY}, GIOPS \cite{GIOPS}, FOAM \cite{FOAM}, and BLUElink OceanMAPS (BLK). 
Details of these forecasting systems are listed in Table~\ref{tab:oceanproduct}. 
For RTOFS, considering that there are only four months of forecasting data from Jan. 2019 to Dec. 2020, thus we do not show the experimental results of RTOFS in this paper. 

\begin{table*}[t]
	\newcommand{\tabincell}[2]{\begin{tabular}{@{}#1@{}}#2\end{tabular}}
	\centering
	\caption{Global ocean forecasting systems compared in this work}
	\begin{tabular}{c|ccccc}
		\toprule
		\tabincell{c}{System} & \tabincell{c}{Horizontal \\resolution} & \tabincell{c}{Number of \\vertical levels} & \tabincell{c}{OGCM} & \tabincell{c}{Data assimilation} &  \tabincell{c}{Forecast range} \\
		\midrule
		PSY4 & $1/12^{\circ}$ & 50 & NEMO 3.1 & SAM 2, 3DVAR & 10 days \\
		FOAM   & $1/4^{\circ}$  & 75 & NEMO 3.2 & NEMOVAR (3DVAR) & 6 days \\
		OceanMAPS (BLK)  & $1/10^{\circ}$ &  51 & OFAM 3 (MOM 5) & BODAS (EnOI) & 7 days \\
		GIOPS & $1/4^{\circ}$   & 50 & NEMO 3.1   & SAM 2         & 10 days \\
            \textit{XiHe}   & $1/12^{\circ}$ &  23  &  XiHe (AI)     &    /       & 10 days \\ 
		\bottomrule
	\end{tabular}%
	\label{tab:oceanproduct}%
\end{table*}%

\subsubsection{Overall Evaluation}
\begin{figure*}
    \centering
    \includegraphics[width=0.85\textwidth]{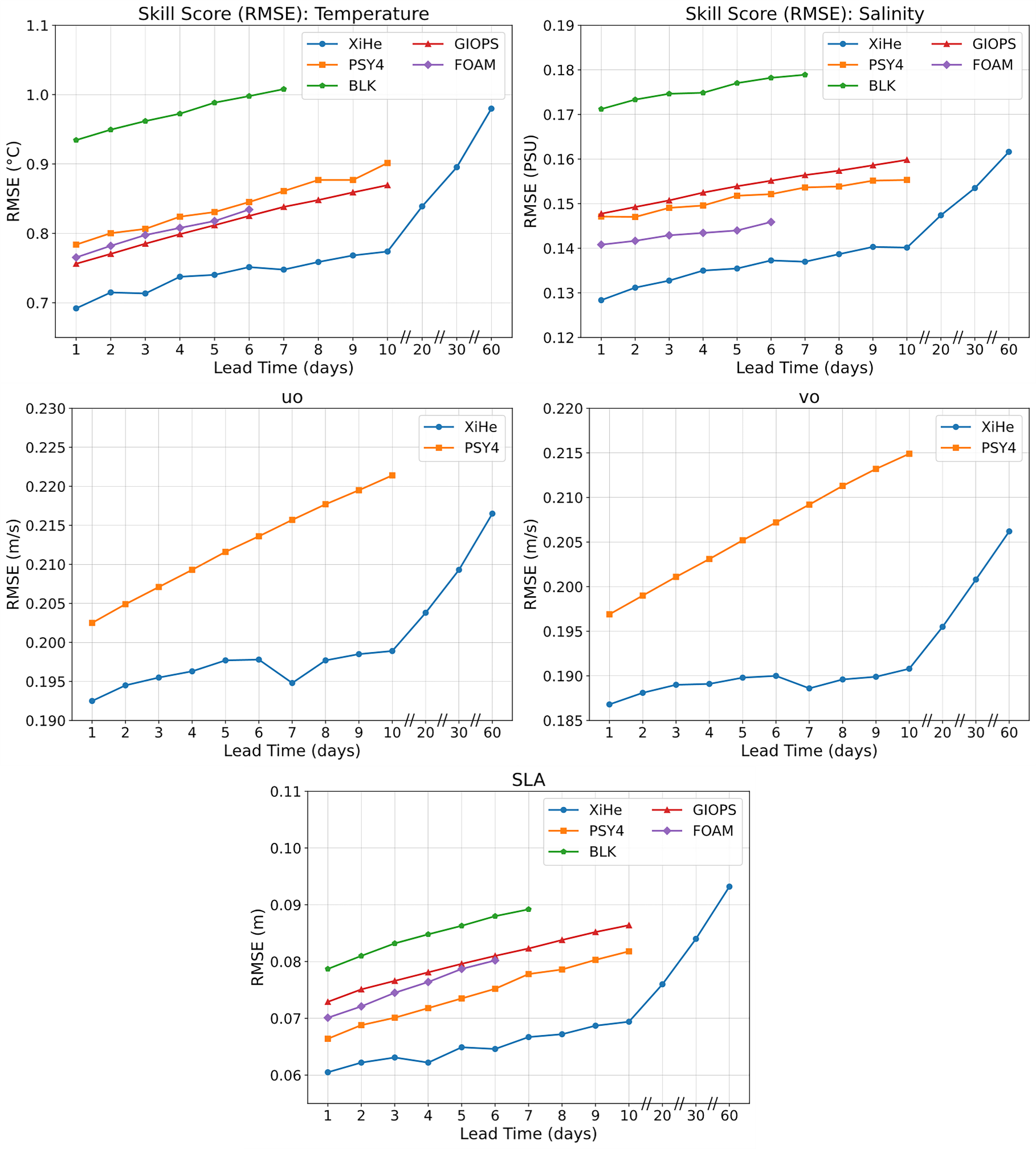}
    \caption{The performance comparison of our data-driven model \textit{XiHe} with the four leading operational numerical GOFSs based on the IV-TT evaluation framework. The x-axis represents the forecast time. The y-axis represents the RMSE (lower is better) of the GOFSs. In addition, for each lead time, the RMSE is averaged from 1st Jan, 2019 to 31st Dec, 2020.
    }
    \label{fig:leading-days}
\end{figure*}

To comprehensively investigate the forecasting capability of \textit{XiHe}, we evaluate the forecast performance over multiple variables, including temperature, salinity, surface current, and SLA, with the lead times ranging from 1 day to 10 days. To further understand the \textit{XiHe} forecasting capabilities, we extended the forecast lead times to 20, 30, and 60 days. 
The experimental results are presented in Fig.~\ref{fig:leading-days}. 
To simplify the display on this figure, the horizontal axis is not scaled linearly after 10 days. 
Note that the PSY4 forecasting system serves as the only baseline in the ocean current evaluation, since only PSY4 provides the ocean current forecast results in the IV-TT framework from Jan. 2019 to Dec. 2020. 
Moreover, BLK and FOAM merely submit forecast results with a maximum duration of 6 days and 7 days to the IV-TT framework, respectively.

For all the evaluated variables including ocean temperature, salinity, 15m-depth current, and SLA, \textit{XiHe} consistently outperforms all the numerical GOFSs. The advantage persists across lead times from 1 day to 10 days.
For ocean temperature, compared to FOAM, PSY4, GIOPS, and BLK in the 144-hour (6-day) lead time forecasting, \textit{XiHe}'s vertical averaged temperature provides an average improvement of 9.94\%, 11.10\%, 8.94\%,  and 24.70\%, respectively. 
Meanwhile, \textit{XiHe}'s vertical averaged salinity provides an average improvement of 5.90\%, 9.77\%, 11.52\%, and 22.98\%, respectively.

Compared with the PSY4 system on the 15m-depth zonal and meridional component of ocean currents, \textit{XiHe} shows a much smaller growing tendency in RMSE. 
The slopes of the error curves can reflect the growth of forecasting error as the lead time increases. 
We calculated the approximate slopes of the error curves for PSY4 and \textit{XiHe} with the lead times from 1 day to 10 days, respectively. 
For the 15m-depth zonal (meridional) component of ocean currents, the slope of PSY4 is 0.00189 (0.00180) $m/(s \cdot day)$, which is about 3.0 times (4.5 times) as large as 0.00064 (0.00040) $m/(s \cdot day)$ of \textit{XiHe}.
This demonstrates that the forecasting error of \textit{XiHe} grows much slower than PSY4, and \textit{XiHe} has the potential for a longer forecast lead time.
It is noteworthy that the 10-day RMSE of \textit{XiHe} is remarkably smaller than the 1-day RMSE of PSY4. 
Furthermore, to assess the forecasting performance of \textit{XiHe} for a longer lead time, we train the models for the lead time of 20 days, 30 days, and 60 days. 
As shown in Fig~\ref{fig:leading-days}, the RMSE values of the 15m-depth zonal component of ocean currents (uo) for the three lead times are 0.2038, 0.2093, and 0.2165 respectively, which is even smaller than the PSY4's 10-day RMSE of 0.2214. 
The RMSE values of the 15m-depth meridional component of ocean currents (vo) for the three lead times are 0.1955, 0.2008, and 0.2062 respectively, which is smaller than the 10-day RMSE of 0.2149 for PSY4. 
This indicates that the accuracy of ocean current forecasting of \textit{XiHe} out to 60 days is even better than that of PSY4 in 10 days.

From the perspective of both the oceanic variables and the lead times, the overall evaluation results firmly demonstrate that the data-driven \textit{XiHe} model has achieved stronger forecast performance than the leading operational numerical GOFSs including PSY4, GIOPS, BLUElink OceanMAPS, and FOAM. 


\begin{figure*}
    \centering
\includegraphics[width=0.85\textwidth]{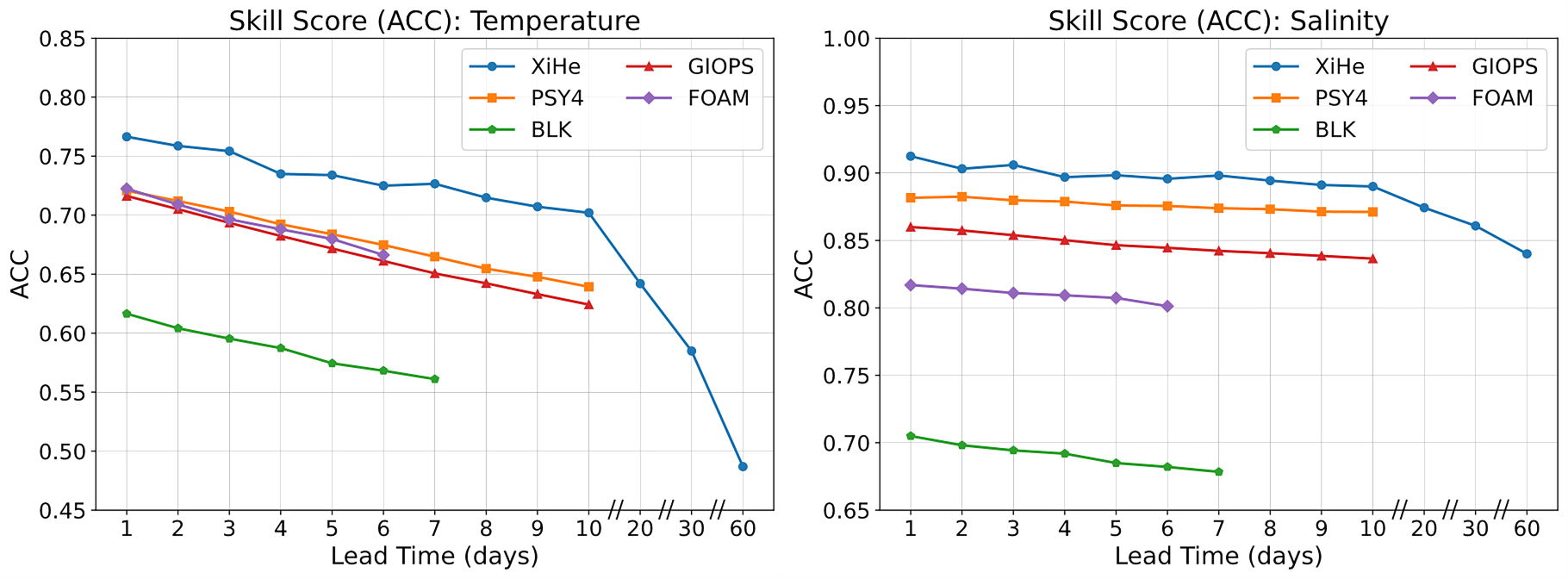}
    \caption{The performance comparison of our data-driven model \textit{XiHe} with the four leading operational numerical GOFSs based on the IV-TT framework. The x-axis represents the forecast time. The y-axis represents the forecasting ACC (higher is better) of the global ocean forecasting systems. In addition, for each lead time, the ACC value is averaged from Jan. 1st, 2019 to Dec. 31st, 2020.}
    \label{fig:ACC}
\end{figure*}

As shown in Fig. \ref{fig:ACC} of the ACC analysis, \textit{XiHe} outperforms the four leading operational numerical GOFSs in terms of temperature and salinity. 
The ACC values of temperature and salinity profiles of \textit{XiHe}'s 20-day forecasting are 0.8743 and 0.6420, which are even better than the 10-day forecasts of other GOFSs. 
This highlights \textit{XiHe}'s robust forecasting capabilities.

\begin{figure*}
    \centering
    \includegraphics[width=\textwidth]{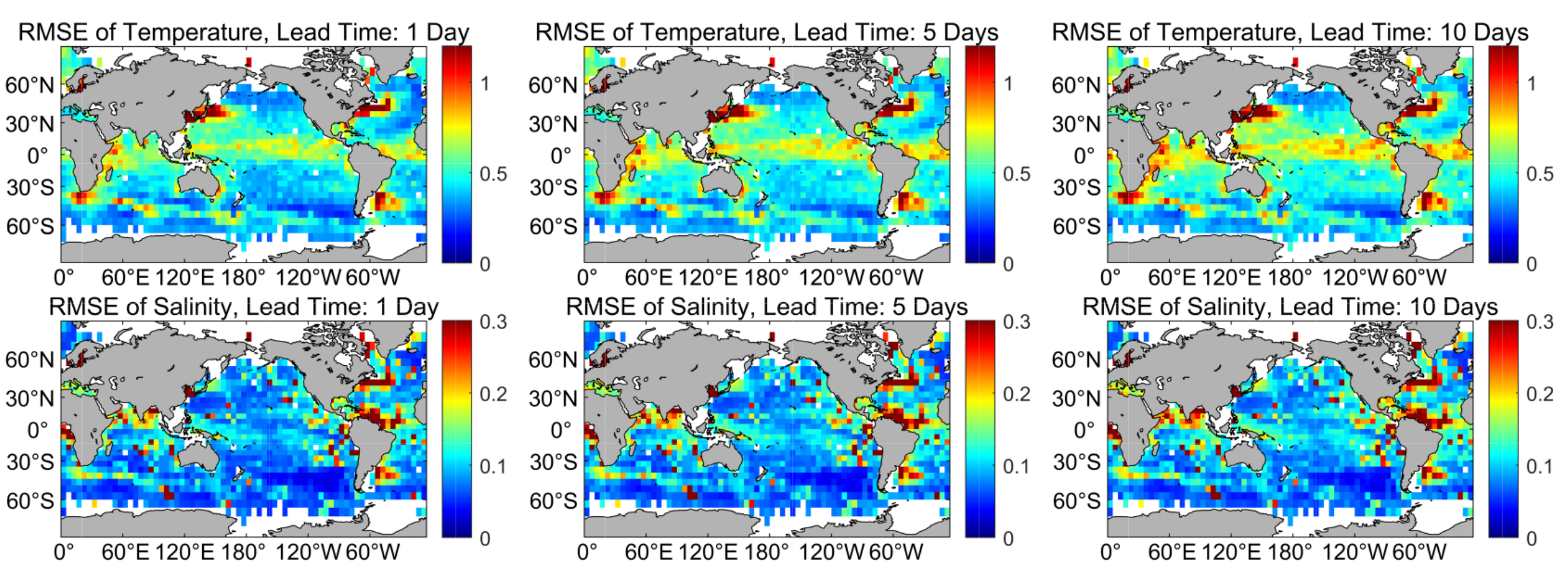}
    \caption{The average RMSEs over vertical depth of the temperature and salinity between observation data and \textit{XiHe} with lead times of 1-day, 5-day, and 10-day calculated on every $5^{\circ} \times 5^{\circ}$ area from Jan. 1st, 2019 to Dec. 31st, 2020. The observation comes from the IV-TT Class 4 framework.}
    \label{fig:global_error}
\end{figure*}

Figure~\ref{fig:global_error} shows the spatial distribution of the average RMSE over vertical depth of the temperature and salinity between \textit{XiHe} and \textit{in situ} observations from the IV-TT Class 4 framework. 
The average RMSE is calculated on every $5^{\circ} \times 5^{\circ}$ area from Jan. 1st, 2019 to Dec. 31st, 2020. 
The forecast results provided by XiHe are consistent with the \textit{in situ} data in most open oceans, since the RMSEs of temperature and salinity are lower than $0.5 ^\circ$C and 0.2 $PSU$.
At the same time, there are some regions where the RMSEs for temperature and salinity are large. These areas include the Agulhas Current, the Kuroshio Extension, the Gulf Stream, the Brazilian Current, the Antarctic Circumpolar Current, and the tropical oceans. 
It is noted that the RMSE of XiHe grows slowly with the increase of lead time.

\begin{figure*}
    \centering
    \includegraphics[width=\textwidth]{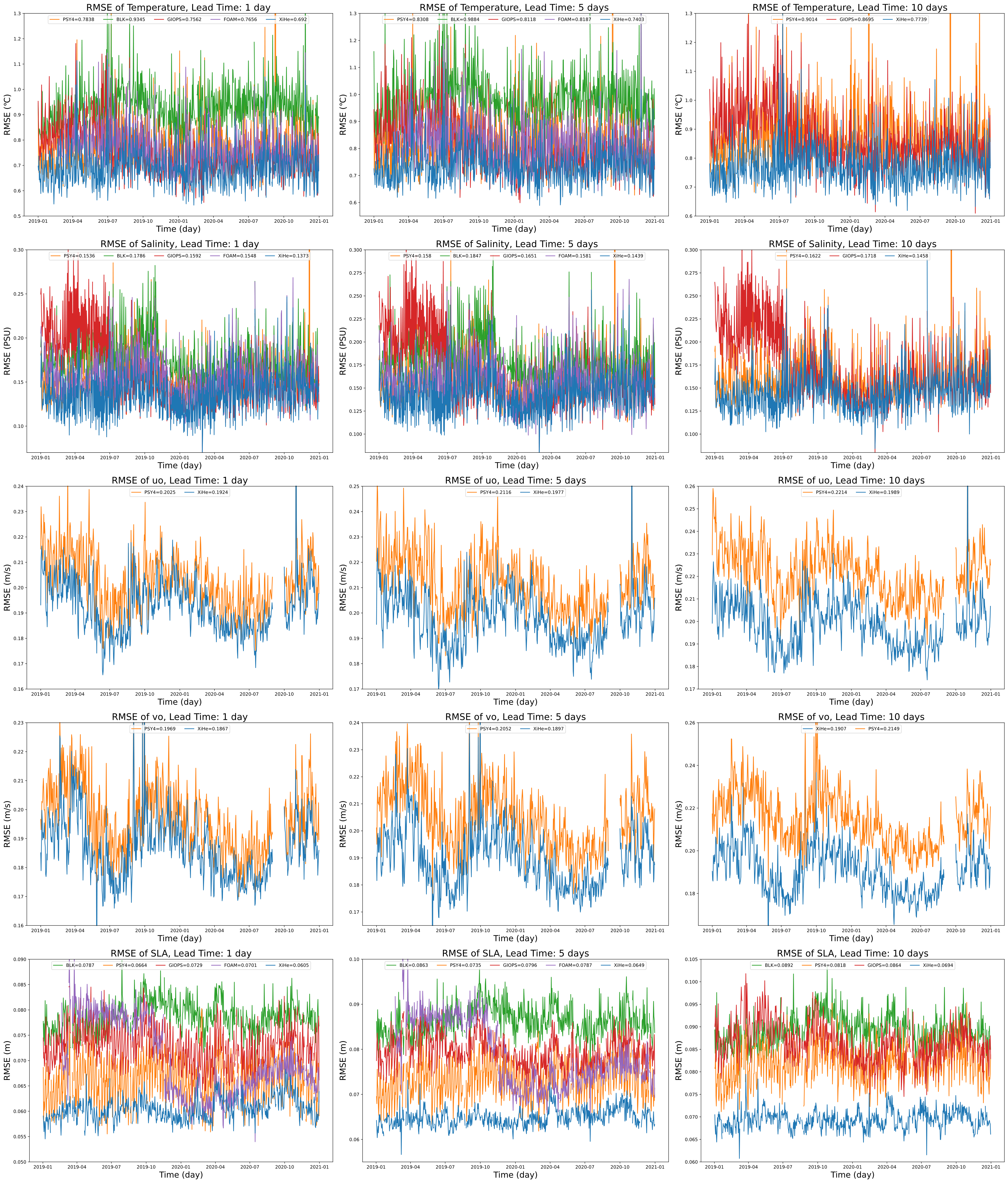}
    \caption{The globally averaged daily RMSE of temperature, salinity, the 15m-depth zonal component of ocean currents (uo), the 15m-depth meridional component of ocean currents (vo),  and the sea level anomaly (SLA) for the forecast lead time of 1 day, 5 days, and 10 days. The x-axis denotes the evaluated time. The y-axis denotes the daily RMSE of the GOFSs' forecast results based on the IV-TT observations.}
    \label{fig:daily-sla-u-v}
\end{figure*}

To be more detailed, we present the daily RMSE of temperature, salinity, uo, vo, and SLA from Jan. 1st, 2019 to Dec. 31st, 2020 in Fig. \ref{fig:daily-sla-u-v} with the lead times of 1-day, 5-day, and 10-day.
For the five variables, \textit{XiHe} exceeds the numerical forecasting systems by an immense margin during the evaluated period. 
Especially for the uo and vo variables, \textit{XiHe} achieves an average improvement of 10.16\% and 11.21\% over PSY4 in the 10-day lead time forecasting, respectively. 
It can be observed that the forecast errors for different variables exhibit certain seasonal variation characteristics. 
For example, with regard to the XiHe and PSY4 models, the forecast errors for ocean currents (uo, vo) are smaller in the boreal summer and larger in the boreal winter. 
As the forecast lead time increases, the seasonal variation in forecast errors becomes more evident.
In addition, it is worth mentioning that the forecast results provided by \textit{XiHe} are more stable than PSY4, exhibiting fewer excessive peak values that would indicate dramatic errors.


\subsubsection{Temperature/Salinity Profiles Evaluation}
\begin{figure*}
    \centering
    \includegraphics[width=\textwidth]{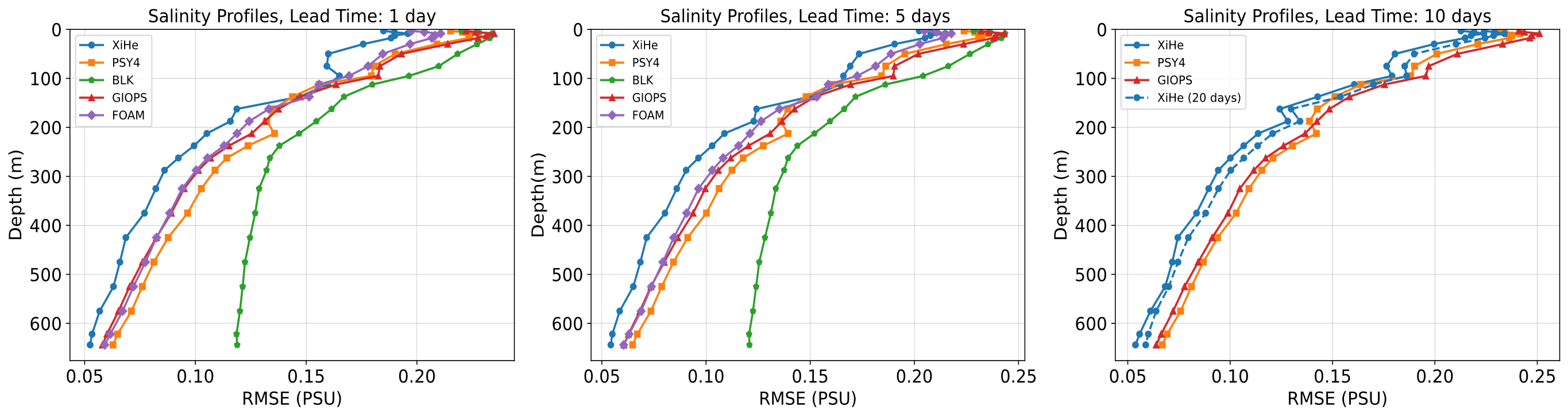}
    \caption{The profiles of salinity RMSE above 650m for different GOFSs. The lead times are 1 day, 5 days and 10 days. For each lead time, the reported RMSE is averaged from Jan. 1st, 2019 to Dec. 31st, 2020.}
    \label{fig:profile-so}
\end{figure*}

\begin{figure*}
    \centering
    \includegraphics[width=\textwidth]{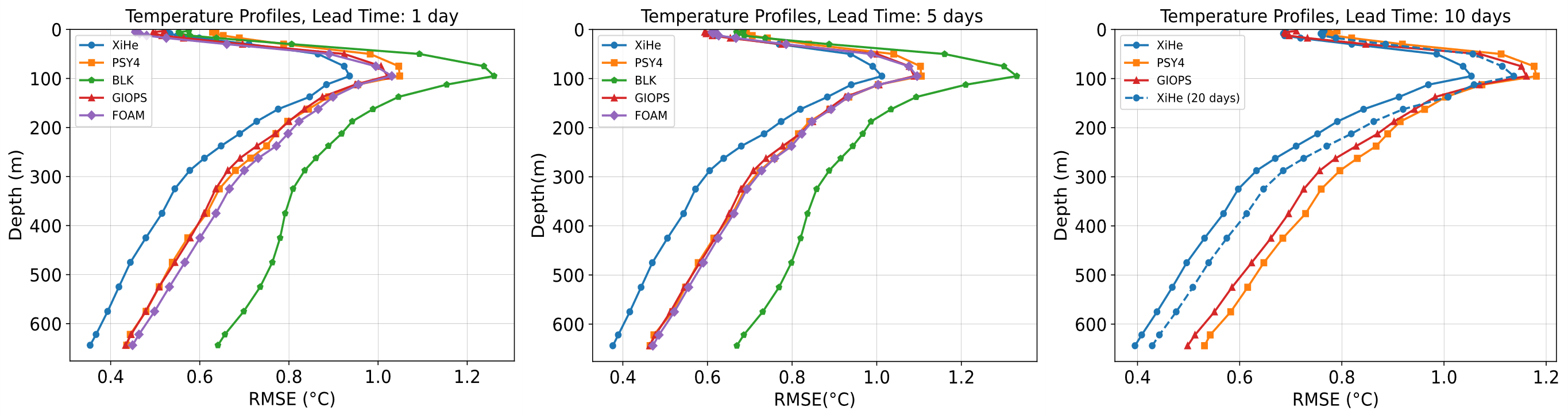}
    \caption{The profiles of temperature RMSE above 650m for different GOFSs. The lead times are 1 day, 5 days and 10 days. For each lead time, the reported RMSE is averaged from Jan. 1st, 2019 to Dec. 31st, 2020.}
    \label{fig:profile-thetao}
\end{figure*}

In order to gain more insights into the forecast errors at different depths, Fig. \ref{fig:profile-so} and Fig. \ref{fig:profile-thetao} show the profiles of the temperature and the salinity RMSE for different GOFSs in depth above 650m. 

As to the depth-wise evaluation results towards the salinity variable in Fig. \ref{fig:profile-so}, \textit{XiHe} outperforms all the compared numerical operational GOFSs almost in the whole depths. And only when the lead time is 1 day and 5 days, for the salinity profiles just from 95m to 137.5m, can FOAM, GIOPS and PSY4 achieve comparable or slightly better performances than \textit{XiHe}. This denotes that \textit{XiHe} performs overall better than the leading GOFSs in accurately forecasting the salinity variable across the entire profile. 

The experimental results in Fig.~\ref{fig:profile-thetao} show the depth-wise evaluation of temperature.
\textit{XiHe} consistently outperforms the four operational numerical GOFSs in terms of temperature variables almost in the whole depths, which is particularly excellent below the depth of 30m. And only when the lead time is 1 day and 5 days, for the temperature profiles just above 30m, can FOAM and GIOPS and achieve slightly better performances than \textit{XiHe}. This denotes that \textit{XiHe} performs overall better than the leading GOFSs in accurately forecasting the temperature variable across the entire profile.


\subsection{Ocean T/S Evaluation with Global Tropical Moored Buoy Array}
In order to better evaluate the performance of \textit{XiHe} on the temperature and salinity variables, we compare the forecast results with the GLORYS12 reanalysis data based on the observations from the Global Tropical Moored Buoy Array, which contains the TAO, RAMA, and PIRATA projects.
The detailed information of the observations is in Section~\ref{sec:EvaDataset}.
We calculate the average RMSE of the forecasting data of \textit{XiHe} and the GLORYS12 reanalysis data based on the observations at a depth ranging from sea surface to 200m over two years (Jan. 1st, 2019 to Dec. 31st, 2020). 

\begin{figure*}
    \centering
    \includegraphics[width=\textwidth, scale=0.8]{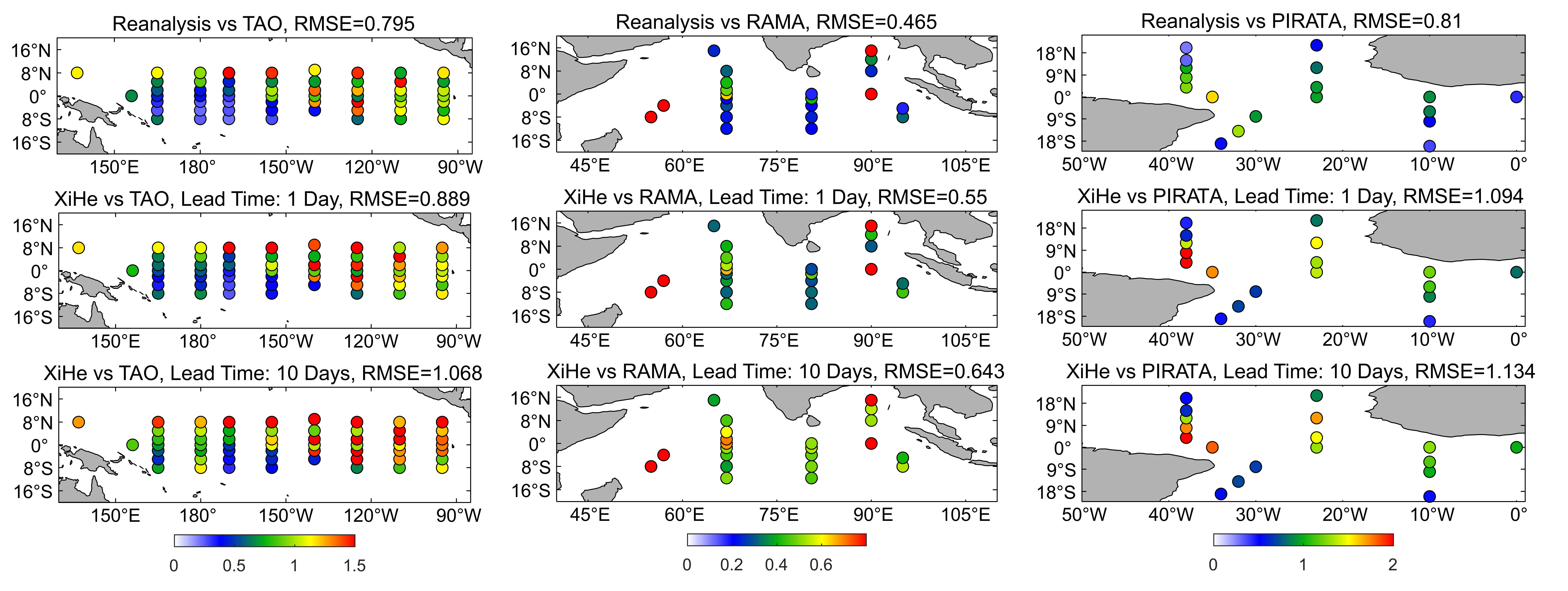}
    \caption{The average RMSE over vertical depth of the temperature between reanalysis data and \textit{XiHe} from Jan. 1st, 2019 to Dec. 31st, 2020. The observation comes from the Global Tropical Moored Buoy Array containing the TAO, RAMA, and PIRATA projects.}
    \label{fig:Buoy_Tem}
\end{figure*}

For the temperature, the average RMSE values of the GLORYS12 reanalysis data in the Atlantic, Pacific, and Indian Ocean regions are 0.810 $^\text{o}$C, 0.795 $^\text{o}$C, and 0.465 $^\text{o}$C as shown in Fig. \ref{fig:Buoy_Tem}. 
In contrast, the average RMSE values of \textit{XiHe} with a lead time of 1 day are 1.094 $^\text{o}$C, 0.889 $^\text{o}$C, and 0.550 $^\text{o}$C in the three regions, respectively. 
Moreover, the average RMSE values of \textit{XiHe} with a lead time of 10 days are 1.134 $^\text{o}$C, 1.068 $^\text{o}$C, and 0.643 $^\text{o}$C, respectively. 
It's worth noting that the forecasting RMSE values of \textit{XiHe} with a lead time of 1 day are close to the GLORYS12 reanalysis data.

\begin{figure*}
    \centering
    \includegraphics[width=\textwidth, scale=0.8]{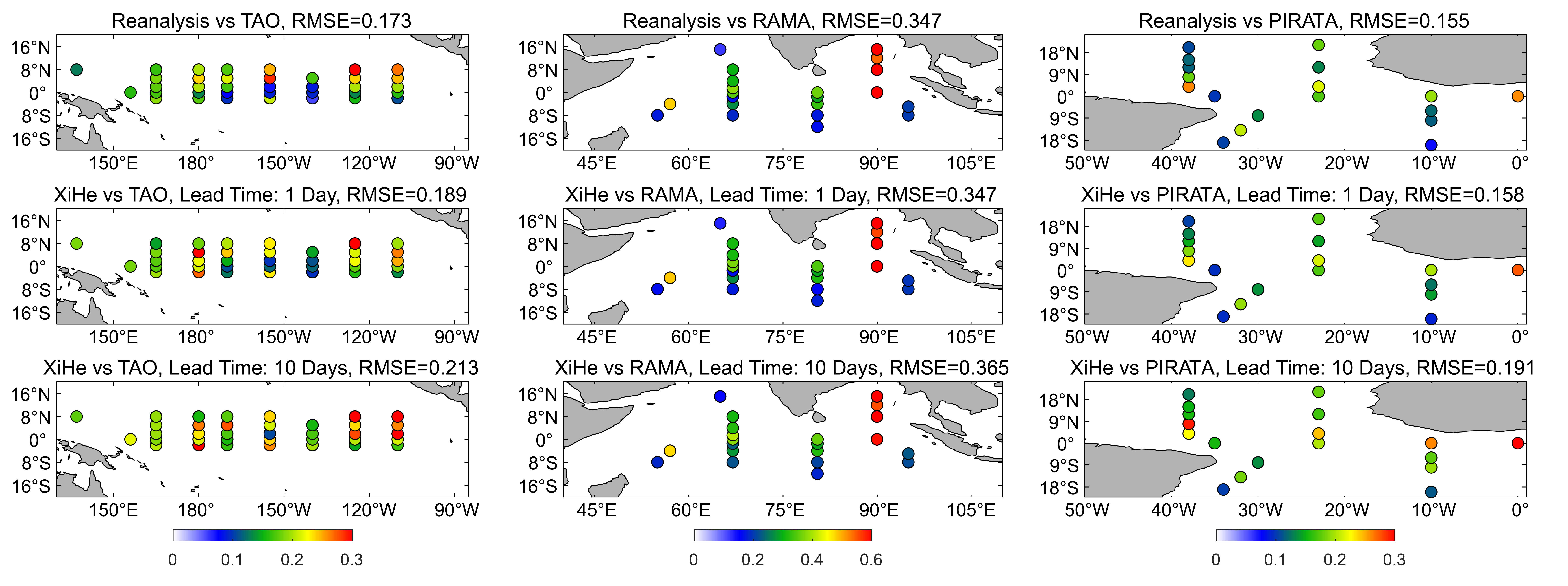}
    \caption{The average RMSE over vertical depth of the salinity between reanalysis data and \textit{XiHe} from Jan. 1st, 2019 to Dec. 31st, 2020. The observation comes from the Global Tropical Moored Buoy Array containing the TAO, RAMA, and PIRATA projects.}
    \label{fig:Buoy_Sal}
\end{figure*}

For the salinity variable, the average RMSE values of the GLORYS12 reanalysis data are 0.155, 0.173, and 0.347 in the three regions respectively as shown in Fig. \ref{fig:Buoy_Sal}. 
In contrast, the average RMSE values of \textit{XiHe} with a lead time of 1 day are 0.158, 0.347, and 0.189 in the three regions.
Moreover, the average RMSE values of \textit{XiHe} with a lead time of 10 days are 0.191, 0.365, and 0.213.
The results show that the forecasting error values of \textit{XiHe} are very close to the GLORYS12 reanalysis data.

\subsection{Large-Scale Ocean Currents Evaluation with Satellite Observation}
\begin{figure*}
\centering
\includegraphics[width=\textwidth]{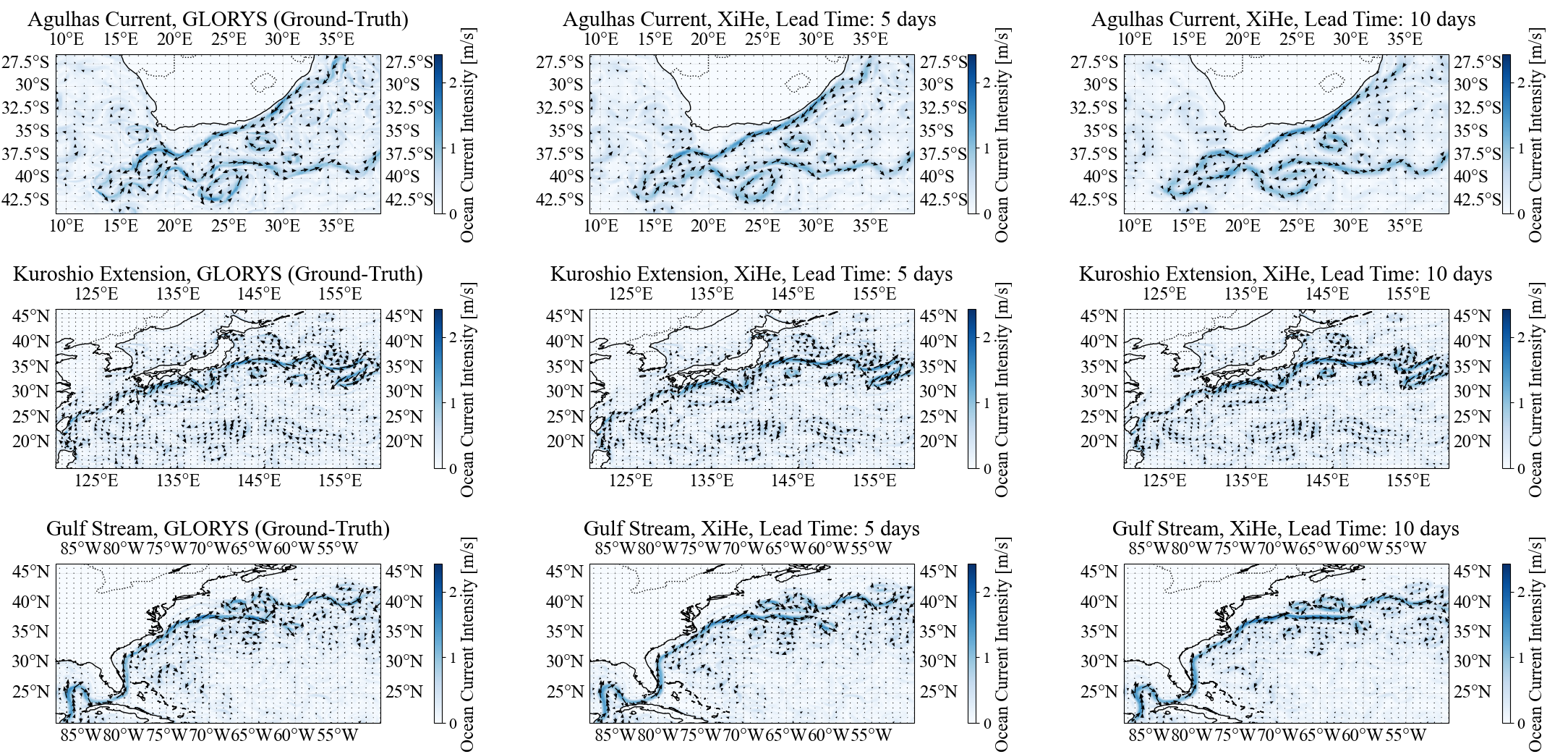}
\caption{Ocean current forecast results of \textit{XiHe} with lead times of 5 and 10 days, as well as the GLORYS12 reanalysis data on Sep. 1st, 2019. The arrows indicate the direction of ocean current and the color represents the intensity of ocean current speed. From top to down, it is Agulhas Current, Kuroshio Current, and North Atlantic Gulf Stream respectively.}
\label{fig:Ocean_Current_Area}
\end{figure*}

One of the most significant and large-scale phenomena in the ocean is ocean circulation. 
Ocean circulations are continuous, directed movements of seawater generated by a variety of forces, including wind, temperature gradients, the Earth's rotation, and differences in salinity. 
It is essential and important to evaluate the ability of \textit{XiHe} to forecast these large-scale ocean phenomena. 
We compare \textit{XiHe} with the GLORYS12 reanalysis data in different forecasting lead times. 
Fig. \ref{fig:Ocean_Current_Area} shows the forecast results of three well-known ocean currents: Agulhas, Kuroshio, and North Atlantic Gulf Stream.
We can find that the forecast results of \textit{XiHe} are very close to the GLORYS12 reanalysis data in ocean current directions and speeds, which verifies the good ability of \textit{XiHe} for large-scale phenomena forecasting.

\begin{figure*}
\centering
\includegraphics[width=0.95\textwidth, scale=0.8]{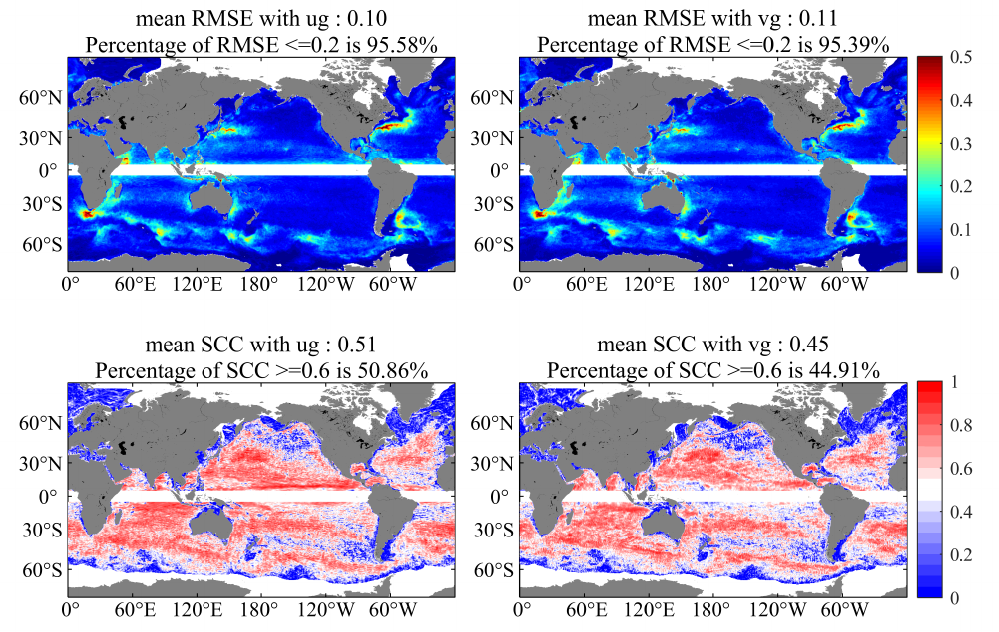}
\caption{RMSEs between \textit{XiHe} surface velocity and geostrophic currents with 5-day forecast lead time in Global Ocean during Jan. 2019 and Dec. 2020. $ug$ represents the zonal geostrophic current, $vg$ represents the meridional geostrophic current, and $CC$ represents the Spearman correlation coefficient. Data have been excluded between 5°S and 5°N, where the geostrophic approximation is not satisfied.}
\label{fig:rmse_with_gc_global}
\end{figure*}

Since the large-scale and mesoscale processes are largely geostrophic or quasi-geostrophic, the surface currents can be evaluated using geostrophic currents derived from altimeter observations. 
The geostrophic current observations for comparison are from CMEMS~\footnote{Global Ocean Gridded L4 Sea Surface Heights And Derived Variables Reprocessed 1993 Ongoing. CMEMS. https://data.marine.copernicus.eu/product/SEALEVEL\_GLO\_PHY\_
L4\_MY\_008\_047/description}.
The evaluation results of the global ocean are shown in Fig.~\ref{fig:rmse_with_gc_global}.
We can find that the RMSE values of geostrophic currents with 5-day forecast lead time are less than 0.2 m/s over 95\% of the global ocean, and the Spearman correlation coefficients (SCC) are larger than 0.6 over 44\% of the global ocean.
The global mean zonal and meridional RMSEs during the two years (from Jan. 2019 to Dec. 2020) are 0.10 m/s and 0.11 m/s, respectively.
The large RMSE mainly appears in the strong current regions, such as the Kuroshio Current, Gulf Stream, Agulhas Current, Southern Ocean, and so on, due to high variability in these regions. 

Table~\ref{tab:GC} also shows the evaluation results of the Indian Ocean, Atlantic Ocean, and Pacific Ocean. 
In all three ocean regions, the percentage of RMSE values below 0.2 m/s exceeds 92.0\% for each region.
And for both $ug$ and $vg$, this percentage in the Pacific Ocean can even reach up to 97.0\%. 
The mean RMSE values of the 5-day forecast in the three oceans are all about 0.10. 
The mean $SCC$ in the Indian Ocean is the largest among the three oceans. 
The percentage of $SCC$ larger than 0.6 appears in the Indian Ocean is the largest, while the small one appears in the Atlantic Ocean.

\begin{table}
\centering
\caption{Mean RMSEs and Spearman correlation coefficients (SCC) of geostrophic currents with 5-day forecast lead time in the three oceans. The geostrophic current observations for comparison are from CMEMS from Jan. 2019 to Dec. 2020. $ug$ represents the zonal geostrophic current, and $vg$ represents the meridional geostrophic current.}
\begin{tabular}{c|c|ccc} 
\hline
\multicolumn{2}{c|}{\textcolor[rgb]{0.2,0.2,0.2}{Statistics}}  & \textcolor[rgb]{0.2,0.2,0.2}{Indian} & \textcolor[rgb]{0.2,0.2,0.2}{Atlantic} & \textcolor[rgb]{0.2,0.2,0.2}{Pacific}  \\ \hline
\multirow{4}{*}{$ug$} & \textcolor[rgb]{0.2,0.2,0.2}{Mean RMSE} & 0.10 & 0.09 & 0.09 \\ 
& \textcolor[rgb]{0.2,0.2,0.2}{Percentage of RMSE $\le$ 0.2m/s} & 92.47\%  & 93.70\% & 97.04\% \\
& \textcolor[rgb]{0.2,0.2,0.2}{Mean SCC}   & 0.60 & 0.49 & 0.57 \\ 
& \textcolor[rgb]{0.2,0.2,0.2}{Percentage of SCC $\ge$ 0.6}   & 61.74\%  & 39.74\%  & 55.62\% \\ \hline
\multirow{4}{*}{$vg$} & \textcolor[rgb]{0.2,0.2,0.2}{Mean RMSE} & 0.11  & 0.10   & 0.09 \\
& \textcolor[rgb]{0.2,0.2,0.2}{Percentage of RMSE $\le$ 0.2m/s} & 92.44\% & 93.06\% & 97.05\% \\
& \textcolor[rgb]{0.2,0.2,0.2}{Mean SCC}   & 0.60 & 0.47 & 0.54 \\
& \textcolor[rgb]{0.2,0.2,0.2}{Percentage of SCC $\ge$ 0.6}   & 58.89\% & 36.16\%  & 47.14\% \\\hline
\end{tabular}
\label{tab:GC}
\end{table}

Above all, the ocean currents forecast data of \textit{XiHe} are consistent well with Altimeter geostrophic current products.  
These results indicate that the \textit{XiHe} model is able to forecast the geostrophic or quasi-geostrophic processes.  


\subsection{Ocean Mesoscale Eddy Evaluation with Satellite Observation}

To evaluate \textit{XiHe}'s ability to portray mesoscale eddy phenomena, we conduct a mesoscale eddy identification experiment focusing on the Northwest Pacific Ocean (15°N-50°N; 110°E-160°E). 
In order to remove small-scale signals and better identify mesoscale eddies, we carried out a simple 11-point smoothing filter process using the Locally Weighted Scatterplot Smoothing method. 
We use satellite SLA data from CMEMS as the approximate truth, since it is widely acknowledged as the authoritative source of satellite data for identifying mesoscale eddies.

\begin{figure*}
\centering
\includegraphics[width=\textwidth]{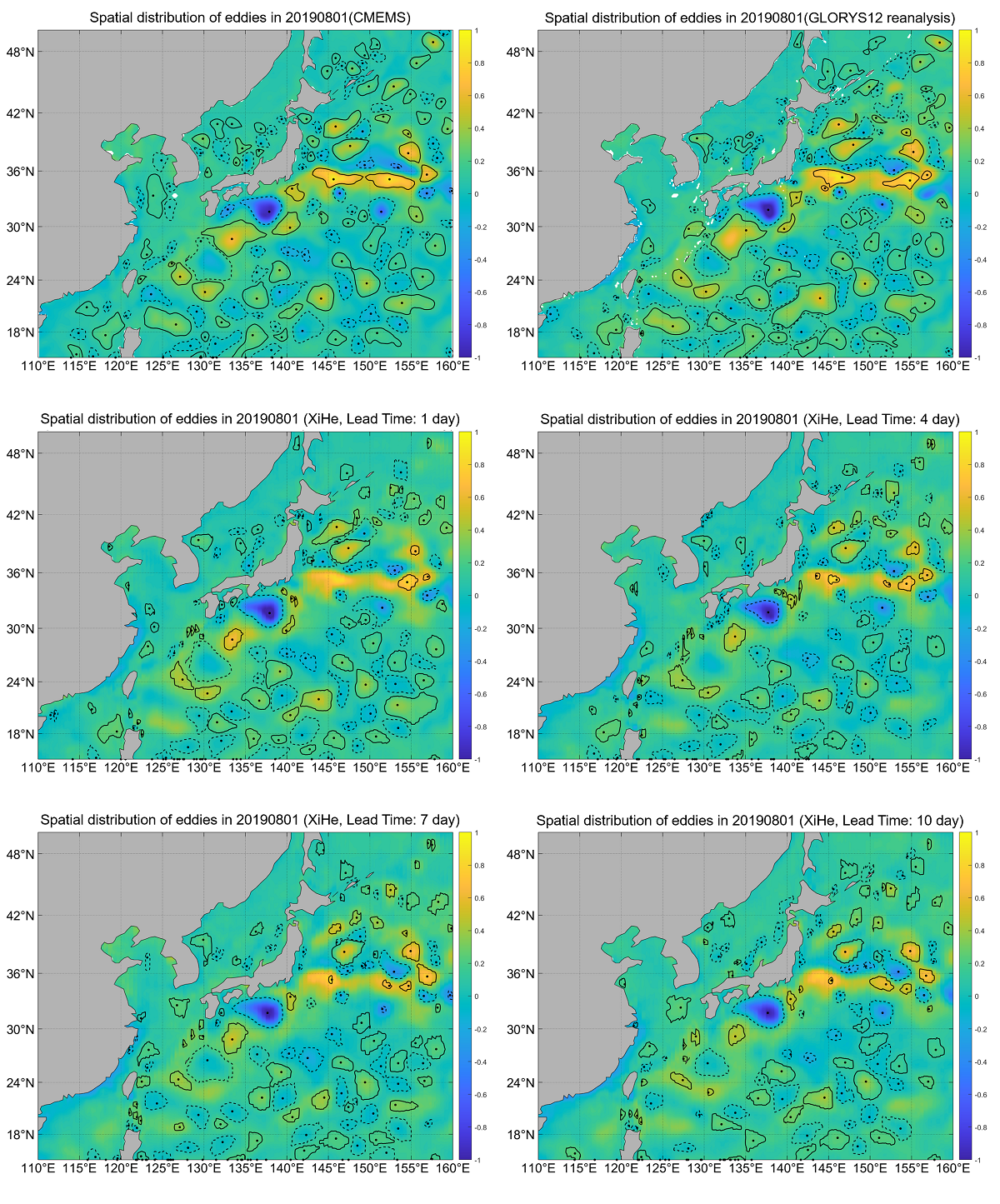}
\caption{Spatial distribution of eddies based on the SLA data obtained from CMEMS, GLORYS12 reanalysis, and \textit{XiHe}'s 1-day, 4-day, 7-day, and 10-day lead time forecast data on Aug. 1, 2019. Dashed lines indicate cyclonic eddies, while solid lines indicate anticyclonic eddies.} 
\label{fig:Eddy_2019}
\end{figure*}

We identified eddies in the area utilizing SLA data obtained from CMEMS, GLORYS12 reanalysis data, and \textit{XiHe}'s forecast data with lead times of 1 day, 4 days, 7 days, and 10 days. 
Taking the eddy identification results on Aug. 1, 2019, as an example, the spatial distribution of the eddies on the day with different datasets is shown in Fig. \ref{fig:Eddy_2019}.
Table~\ref{tab:eddiesNumber} presents the corresponding number of detected eddies from the different datasets and the count of these eddies that overlap with those identified by CMEMS.
There are 71 cyclonic eddies and 67 anticyclonic eddies detected from the SLA data of CMEMS. 
From the GLORYS12 reanalysis data, there are 66 cyclonic eddies and 56 anticyclonic eddies detected, with 42 cyclonic eddies and 41 anticyclonic eddies overlapping with the CMEMS data. 
For \textit{XiHe}'s 1-day lead time forecasting data, there are 57 cyclonic eddies and 65 anticyclonic eddies detected, with 39 cyclonic eddies and 40 anticyclonic eddies overlapping with the CMEMS data.
The results show that the eddies detected from the reanalysis data and \textit{XiHe}'s forecasting data are close to each other, and also close to those detected from the data of CMEMS. 
As expected, the overlapping number of mesoscale eddies detected from \textit{XiHe}'s forecasting data decreases as the forecast lead days increase.
This indicates that \textit{XiHe} has a strong ability to resolve mesoscale eddies.

\begin{table}[]
\caption{Number of detected eddies from different datasets and overlapping eddies with the CMEMS data on Aug. 01, 2019. The data from \textit{XiHe} here includes the forecast results with different lead times, specifically at 1-day, 4-day, 7-day, and 10-day intervals.}
\begin{tabular}{c|cc|cc}
\hline
\multirow{2}{*}{Datasets} & \multicolumn{2}{c|}{Cyclonic Eddies}  & \multicolumn{2}{c}{Anticyclonic Eddies} \\ \cline{2-5} 
 & \multicolumn{1}{c|}{Detected} & Overlapping & \multicolumn{1}{c|}{Detected} & Overlapping \\ \midrule
CMEMS & \multicolumn{1}{c|}{71} & /  & \multicolumn{1}{c|}{67} & / \\ 
GLORYS12 & \multicolumn{1}{c|}{66} & 42 & \multicolumn{1}{c|}{56} & 41 \\ 
\textit{XiHe} (1-day)  & \multicolumn{1}{c|}{57} & 39 & \multicolumn{1}{c|}{65} & 40 \\ 
\textit{XiHe} (4-day) & \multicolumn{1}{c|}{56} & 37 & \multicolumn{1}{c|}{60} & 35 \\ 
\textit{XiHe} (7-day) & \multicolumn{1}{c|}{56} & 34 & \multicolumn{1}{c|}{65} & 30 \\ 
\textit{XiHe} (10-day) & \multicolumn{1}{c|}{47} & 30 & \multicolumn{1}{c|}{62} & 29 \\ \hline
\end{tabular}
\label{tab:eddiesNumber}
\end{table} 

\begin{figure*}
\centering
\includegraphics[width=\textwidth]{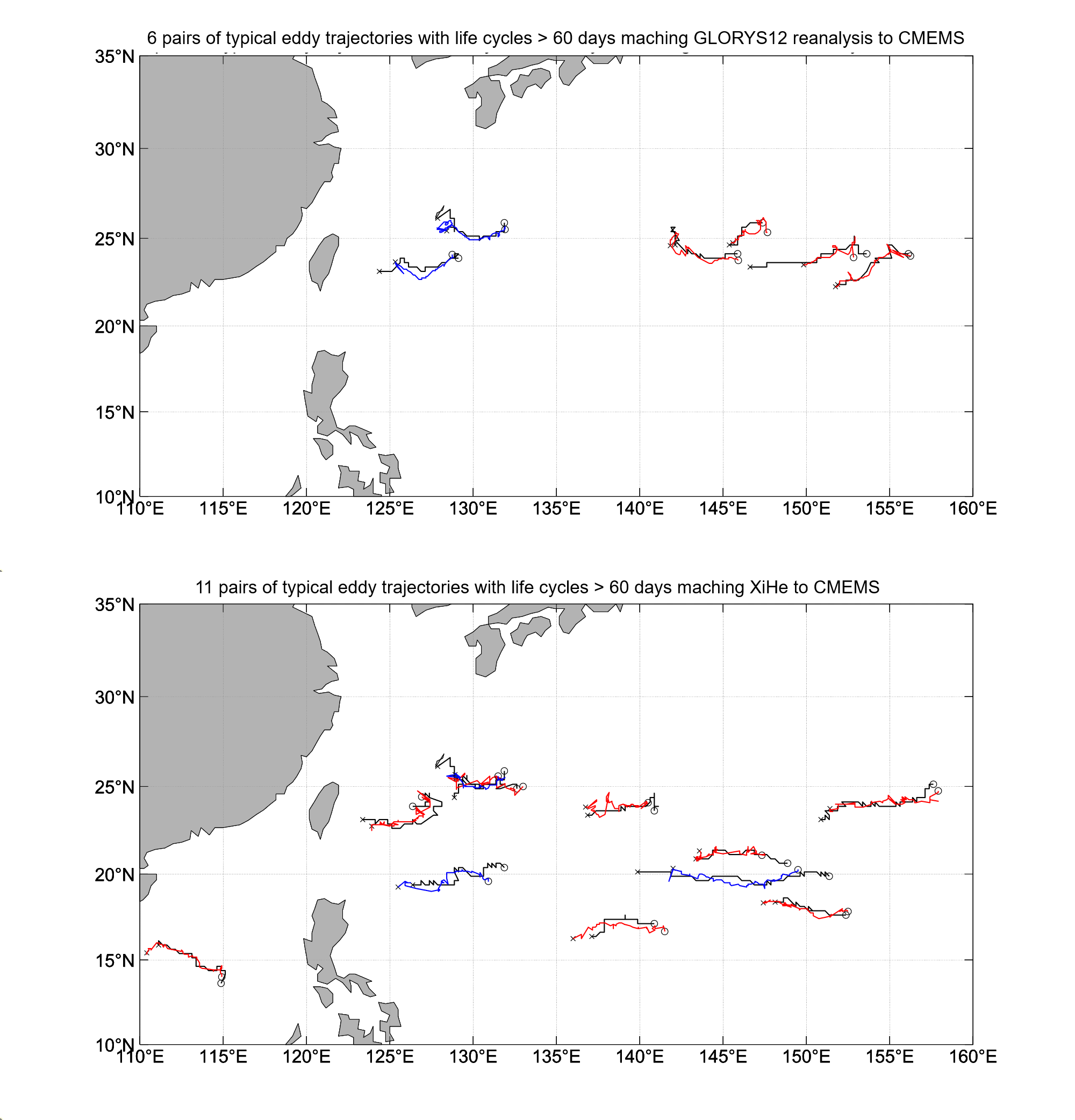}
\caption{6 pairs of overlapping typical eddy trajectories with lifetimes exceeding 60 days matched between GLORYS12 reanalysis and CMEMS, while 11 pairs of overlapping typical eddy trajectories matched between \textit{XiHe} and CMEMS. Blue color indicates cyclonic eddies, while red color indicates anticyclonic eddies.} 
\label{fig:Eddy}
\end{figure*}

As we all know, forecasting eddy trajectory is a big challenge. In order to further test \textit{XiHe}'s capability in forecasting the mesoscale eddy trajectory, we analyze the number of overlapping eddy trajectories of \textit{XiHe}'s forecasting data and GLORYS12's reanalysis data with CMEMS observations. 
We select the eddy trajectories with lifetimes larger than 60 days that appeared in both CMEMS data and GLORYS12 reanalysis data from Jan. 1, 2019, to Dec. 31, 2020. 
We only find 6 overlapping trajectories as shown in Fig.~\ref{fig:Eddy}. 
Following the same methodology, we find 11 pairs of overlapping trajectories that appeared in both CMEMS data and \textit{XiHe}'s forecasting data. 
 
It shows the good ability of \textit{XiHe} in forecasting the mesoscale eddy trajectory. 
\textit{XiHe} can promise a stronger eddy forecasting capacity with the quality improvement of the reanalysis data in the future.


\section{Conclusion and Discussion} \label{sec:conclusion}
In this paper, we propose the first data-driven $1/12^{\circ}$ resolution global ocean eddy-resolving forecasting model \textit{XiHe}, which is trained on 25-year ocean reanalysis data. 
An ocean-land masking mechanism is proposed to make \textit{XiHe} focus more on the global ocean information by excluding the impact of land on the model training. 
Furthermore, a novel ocean-specific block containing both local and global SIE modules is designed to capture the inherent oceanic spatial information. 
\textit{XiHe} achieves stronger forecast performance in all testing variables than existing leading operational numerical GOFSs including PSY4, GIOPS, BLUElink OceanMAPS, and FOAM with thousands of times speeding up. 
Notably, the 60-day ocean current forecasting of \textit{XiHe} is even better than that of PSY4's 10-day forecasting. 
The ACC values of temperature and salinity profiles of \textit{XiHe}'s 20-day forecast are 0.6177 and 0.5260, which are better than both PSY4's and GIOPS's 10-day forecasts. 
In addition, \textit{XiHe} is able to forecast the multi-scale ocean processes, including the large-scale circulation and the mesoscale eddies.

In the future, it would be interesting to further investigate whether data-driven models and numerical GOFSs can be combined and mutually enhance each other.
On one hand, data-driven models can benefit from larger and more accurate reanalysis datasets, which are generated by leveraging numerical GOFSs. 
On the other hand, the initial fields of the data-driven models still mainly come from the data assimilation methods used in numerical forecasting. How to combine data assimilation methods and large models to form better initial fields and forecasting results is a worthwhile endeavour.



%



\ifCLASSOPTIONcompsoc
  \section*{Acknowledgments}
\else
  \section*{Acknowledgment}
\fi

X.W., H.Z.W., G.H.W, and P.Q.W. were supported by National Natural Science Foundation of China (Grant No. 62372460, 42276205, 42288101, 42030405, 42306040). W.M.Z. was supported by National Key R\&D Program of China (Grant No. 2021YFC3101500). H.Z.W. and X.W. were supported by Hunan Provincial Natural Science Foundation of China (2023JJ10053, 2024JJ4042). K.J.R. were supported by the Science and Technology Innovation Program of Hunan Province (2022RC3070). The authors would like to thank the Intercomparison and Validation Task Team of the OceanPredict Community for providing the observations and the forecasts of the operational oceanography forecasting systems.

\ifCLASSOPTIONcaptionsoff
  \newpage
\fi



%
\bibliographystyle{IEEEtran}
\bibliography{IEEEabrv,main}






\
\end{document}